\documentclass[aps,twocolumn,floats,prd,nofootinbib,superscriptaddress,10pt]{revtex4-2}

\def\be{\begin{equation}}
\def\ee{\end{equation}}

\def\21cmf{\texttt{21cmFAST}}

\makeatletter
\def\hlinewd#1{%
\noalign{\ifnum0=`}\fi\hrule \@height #1 \futurelet
\reserved@a\@xhline}
\makeatother

\renewcommand{\arraystretch}{1.}

\usepackage{pifont}

\usepackage{amsmath}
\usepackage{bm}
\usepackage{amssymb}
\usepackage{graphicx}
\usepackage{multirow}
\usepackage{makecell}
\usepackage{hyperref}
\usepackage{adjustbox}
\usepackage{tabularx}
\hypersetup{
	colorlinks=true,
	linkcolor=red,
	filecolor=magenta,      
	urlcolor=blue,
}
\usepackage[normalem]{ulem}
\usepackage[usenames,dvipsnames]{xcolor} 
\usepackage{comment}
\usepackage{makecell}
\usepackage{subfigure}
\usepackage{placeins}
\usepackage{booktabs}
\usepackage{array}

\newcolumntype{C}{>{\centering\arraybackslash}X}

\begin{document}

\title{Towards a multi-tracer neutrino mass measurement with line-intensity mapping}

\author{Gali Shmueli}
\email{shmugal@post.bgu.ac.il}
\affiliation{Department of Physics, Ben-Gurion University Be'er Sheva 84105, Israel}
\author{Sarah Libanore}
\email{libanore@bgu.ac.il}
\affiliation{Department of Physics, Ben-Gurion University Be'er Sheva 84105, Israel}
\author{Ely D. Kovetz}
\email{kovetz@bgu.ac.il}
\affiliation{Department of Physics, Ben-Gurion University Be'er Sheva 84105, Israel}

\begin{abstract}
Accurately determining neutrino masses is a main objective of contemporary cosmology. Since massive neutrinos affect structure formation and evolution, probes of large scale structure are sensitive to the sum of their masses. In this work, we explore future constraints on $\sum m_\nu$ utilizing line-intensity mapping (LIM) as a promising emerging probe of the density of our Universe, focusing on the fine-structure [CII] line as an example, and compare these constraints with those derived from traditional galaxy surveys. Additionally, we perform a multi-tracer analysis using velocity tomography via the kinetic Sunyaev-Zeldovich and  moving lens effects to reconstruct the three-dimensional velocity field. 
Our forecasts indicate that the next-generation AtLAST detector by itself can achieve $\sigma_{\Sigma m_\nu} \sim 50$ meV sensitivity. Velocity tomography will further improve these constraints by 4\%. 
Incorporating forecasts for CMB-S4 and DESI-BAO in a comprehensive multi-tracer analysis, while setting a prior on the optical depth to reionization $\tau$ derived using 21-cm forecasted observations, to break degeneracies, we find that a $\gtrsim5\sigma$ detection of $\sum m_\nu\!\sim\! 60$ meV, under the normal hierarchy, is within reach with LIM. Even without a $\tau$ prior, our combined forecast reaches $\sigma_{\Sigma m_\nu} \!\sim\! 18$ meV.
\end{abstract}

\maketitle

\section{Introduction}\label{sec:intro}

As one of the least understood components of the Standard Model of particle physics, neutrinos have profound implications for both cosmology and astrophysics.
The quest to understand their fundamental properties remains one of the pivotal challenges in particle physics: so far, neutrino oscillation experiments have revealed that these particles have mass and obey one among three possible hierarchies: normal, inverted, or degenerate \cite{Super-Kamiokande:1998kpq,Super-Kamiokande:2001bfk,Super-Kamiokande:2002ujc,Soudan2:2003qqa,Drexlin:2003fc,KamLAND:2004mhv,Super-Kamiokande:2005mbp,Messier:2006yg}. The current status of the experiments, however, does not allow a direct measurement of the mass of each species.

Due to their non-zero mass, neutrinos contribute to the total energy density of the Universe and affect the cosmic expansion rate and the evolution of cosmic structures~\cite{Eisenstein:1997jh, Lesgourgues:2006nd, Brandbyge:2010ge, Marulli:2011he, Castorina:2015bma,Gawiser:2000az}. 
Their signatures can be found in the power spectrum of the cosmic microwave background (CMB) fluctuations, as well as in the statistical analysis of large scale structure (LSS). Potentially, this allows us to use cosmological datasets to probe the sum of neutrino masses, $\sum m_\nu$, and solve the hierarchy problem. Upper bounds on this parameter have already been set both from Planck~\cite{Planck:2018nkj, Planck:2018vyg} and from the analysis of baryon acoustic oscillations (BAO) in DESI 1-year data~\cite{DESI:2024mwx, Allali:2024aiv}. In particular, combining Planck 2018~\cite{Planck:2018nkj,Planck:2018vyg}, ACT lensing~\cite{ACT:2022,ACT:2023} and DESI, Refs.~\cite{DESI:2024a,DESI:2024b} constrain $\sum m_\nu<0.072\,{\rm eV}$ at 95\% CL, this result being however influenced by the lensing anomaly in Planck 2018 data and by the $z=0.7$ bin in DESI analysis, which seems to be in tension with Planck results~\cite{NaredoTuero:2024}. Moreover, the mild preference for a negative value of $\sum m_\nu$ by the data, pointing to an excess of clustering, has led to suggested explanations via effects beyond the standard cosmological model on neutrino physics, e.g.,~cooling or evolving $\sum m_\nu$ in cosmic time, or dynamical/mirage dark energy models ~\cite{Craig:2024tky,
Green:2024xbb,Elbers:2024sha}. 

Robust measurements, however, remain challenging and elusive: to reach a significant detection of $\sum m_\nu$, therefore, one should exploit as many sources of information as possible. This motivates a multi-tracer approach as the best way to reach a solid, reliable result. 

So far, the constraining power of galaxy surveys on neutrino masses has  been explored widely (see e.g.~\cite{Cerbolini:2013uya, Swanson:2010sk}); on the other hand, the field of line-intensity mapping has only recently started to gain attention. 
Line-intensity mapping (LIM,~see Refs.~\cite{Kovetz:2017agg,Bernal:2022jap} and within for review) is an emerging approach to survey the Universe, using relatively low-aperture instruments to collect the total spectral-line emission from galaxies and the intergalactic medium (IGM). Unlike galaxy surveys, LIM does not require resolved high-significance detections, but uses all incoming photons from any source within the field of view. LIM  offers a unique opportunity to probe redshifts well beyond the reach of other cosmological observations, over large patches of the sky, accessing regimes that cannot be explored otherwise~(see,~e.g., Refs.~\cite{Schaan:2021hhy,Bernal:2020lkd,Bernal:2022wsu,Libanore:2022ntl,Adi:2023qdf,Fronenberg:2023qtw}). 

Line emission associated with process of star formation inside cooling molecular clouds, e.g.,~{CO, [CII], [OIII]...}, are currently observed in the spectra of low redshift galaxies and targeted by preliminary LIM experiments~\cite{Stutzer:2024rps,CONCERTO:2020ahk}, while current and next-generation detectors (e.g.,~COMAP-EoR~\cite{Cleary:2021dsp}, CCAT-prime~\cite{CCAT-Prime:2021lly}, SPHEREx~\cite{SPHEREx:2014bgr}) foresee their detection up to $z\sim 8$. The high frequency resolution of these detectors will also enable us to perform high-resolution redshift tomography, providing a good reconstruction of the growth of structure along the line of sight (LoS).
Mapping the intensity fluctuations of different lines, it will be possible to access a large cosmic volume: under these conditions, we should expect LIM to be capable of probing neutrino signatures on a variety of scales, hence potentially measuring $\sum m_\nu$ with high sensitivity~\cite{MoradinezhadDizgah:2021upg,Karkare:2022bai}.

At the same time, interferometers such as HERA~\cite{DeBoer:2016tnn} and SKA~\cite{Ghara:2016dva} are currently being built to achieve the measurement of the hydrogen 21-cm line from reionization and cosmic dawn.\footnote{The modeling of the 21-cm signal, see e.g.,~Refs.~\cite{Furlanetto:2006jb,Pritchard:2011xb}, is significantly different from the one adopted for star-forming related lines. For this reason, when referring to LIM studies throughout this paper, we always implicitly refer to surveys that probe star-forming lines, using [CII] as our main case study.} More than serving as a LSS tracer, the 21-cm signal will also provide information on the conditions of the IGM at $z\sim 6-30$: among other interesting results, this will lead to a clear measurement of the optical depth at reionization $\tau$, whose value is mainly determined by the abundance of free electrons during these stages of cosmic evolution~\cite{Sailer:2022vqx}. In the context of neutrino studies, it has been shown in Refs.~\cite{Liu:2015txa, Shmueli:2023box} that the improved measurement of $\tau$ that e.g.,~HERA will provide, will be crucial to break internal degeneracies in CMB data, potentially allowing a $5\sigma$ detection of $\sum m_\nu$. 

In addition, the signatures of massive neutrinos on the growth of structure can be probed using velocity tomography. 
The cosmological velocity field can be reconstructed via cross-correlating CMB maps with some LSS tracer, specifically harnessing the power of CMB secondary temperature anisotropies, namely the kinetic Sunyaev-Zel'dovich\footnote{Thermal Sunyaev-Zel'dovich effect~\cite{Zeldovich:1969ff,Sunyaev:1972eq,Birkinshaw:1984} is as well relevant in this context (see,~e.g., Refs.~\cite{Hill:2013baa,Roncarelli:2014jla,Mccarthy:2017yqf,Bolliet:2019zuz}).} and moving lens effects.

The kinetic Sunyaev-Zel’dovich (kSZ,~\cite{Zeldovich:1969ff,Zeldovich:1969sb,Sunyaev:1970er,Sunyaev:1980vz,Sazonov:1999zp}) effect is a secondary CMB anisotropy induced by the scattering of CMB photons off intervening electrons in the IGM of galaxy clusters, resulting in a measurable shift in the CMB temperature fluctuation spectrum. 
The kSZ effect has  been detected at a few-sigma by cross-correlating Planck CMB measurements with galaxy-cluster observations in the local Universe~\cite{Hand:2012ui,Planck:2015ywj,ACTPol:2015teu,DES:2016umt,Hill:2016dta,DeBernardis:2016pdv}, and the prospect of its measurement with future experiments, e.g.,~CMB Stage 4 (CMB-S4,~\cite{CMB-S4:2016ple}) or the Simons Observatory~(SO,~\cite{SimonsObservatory:2018koc}), is therefore guaranteed.
The kSZ effect encodes information on the radial LoS velocity, which can be accessed through cross-correlation of the CMB maps with LSS tracers. This potentially provides a tool to constrain cosmology~\cite{Smith:2018bpn,Munchmeyer:2018eey,Bhattacharya:2006ke,Mueller:2014dba,Vanzan:2023cze,Bhattacharya:2007sk,Ma:2013taq,Mueller:2014nsa,Bianchini:2015iaa,Krywonos:2024mpb}, but may be strongly dependent on astrophysical properties
e.g.,~the state of the ionized IGM~\cite{Hernandez-Monteagudo:2015cfa,McGaugh:2007fj} and of the intra-cluster gas~\cite{Haehnelt:1995dg,Diaferio:2004gk}, or the cosmic evolution of helium reionization~\cite{Caliskan:2023yov}. 

To break the internal degeneracies in the reconstruction of the velocity field, a possible solution is to access its transverse component by relying on the moving-lens effect (ML, \cite{Birkinshaw:1983,Tuluie:1995pg,
Aghanim:1998ux,Cooray:2002ee}). This is due to the time-evolving gravitational potential that CMB photons experience while travelling across clusters in transverse motion with respect to the LoS. ML induces additional temperature anisotropies in the CMB via non-linear Integrated Sachs-Wolfe (ISW,~\cite{Sachs:1967er}) effect.
Despite the signal being weak for a single lens, Ref.~\cite{Hotinli:2018yyc} showed that it is possible to gain a relevant statistical signal when the ML effect is sourced by objects having a common bulk motion. Its detection is more challenging than the kSZ one~\cite{Tuluie:1995ut,
Aghanim:1998ux,Cooray:2002ee,Hotinli:2023ywh}, and for this reason its applications are yet largely unexplored; despite this, in principle ML can be used to reconstruct the velocity field on large scales, in order to obtain information on the underlying cosmological properties~\cite{Hotinli:2021hih}.

So far, the kSZ and ML effects in the context of cosmology have been mainly exploited using the cross correlation between galaxies and the CMB (see,~e.g., Refs.~\cite{Smith:2018bpn,Munchmeyer:2018eey,Hotinli:2021hih,Tishue:2024}). As first suggested by Ref.~\cite{Sato-Polito:2020cil}, instead, we are interested in exploring how to access them using LIM surveys. Advantages in this case can arise because of the higher redshifts probed, as well as from the possibility of performing high-resolution tomography. Throughout this paper, we will further explore the comparison between the two LSS tracers in order to constrain $\sum m_\nu$, and we will  investigate the various regimes where one proves to be more advantageous than the other. 

The paper is structured as follows. We first outline the experiments that we consider as targets for our analysis in Sec.~\ref{sec:experiments}; as for LIM, we focus on the study of the [CII] line, but our results can be straightforwardly extended to other cases. In Sec.~\ref{sec:methods}, we detail the main observables used in our analysis and explain the methodology used in this work. We then describe our analysis and results in Sec.~\ref{sec:analyis}, and in Sec.~\ref{sec:summary} we summarize our main findings.

\section{Experiments}
\label{sec:experiments}

\subsection{LIM}
Among the different lines that are observed in LIM studies, the bright fine-structure [CII] line at $\nu = 158\,{\rm \mu m}$ is one of the most widely discussed in the literature. 
Produced by photo-dissociation in molecular gas clouds~\cite{Stacey:2010ps}, it is a well-known tracer of star formation history~\cite{Suginohara:1998ti,DeLooze:2011uw,Herrera-Camus:2014qba} and can potentially be used to probe the redshift range $ z\lesssim 8$. 
Current and upcoming instruments that have [CII] as their main target line include, e.g.,~EXCLAIM~\cite{Ade:2019ril},~TIM~\cite{Vieira:2020},~CCAT-prime~\cite{Choi:2019rrt},~TIME~\cite{Sun:2020mco} and AtLAST~\cite{Klaassen:2019,Booth:2024jkg,Booth:2024oem}.

In this work, through the analysis in Sec.~\ref{sec:analyis}, we want to understand how constraints on $\sum m_\nu$ will improve  as [CII] surveys improve their sensitivity, either when relying on the LIM survey alone, or in combination with the CMB to perform velocity tomography reconstruction. To characterize a LIM survey, we rely on these parameters: 
\begin{itemize}
\item $\Omega_{\rm sky}$: the sky area observed by the survey (in deg$^2$);
\item $t_{\rm obs}$: the total observational time (in $\rm hr$);
\item $\theta_{\rm FWHM}$: the angular resolution of the detector (in arcsec);
\item $\Delta \nu$: the observed bandwidth (in GHz), that sets the observed redshift range, via $\nu_{\rm obs}=\nu(1+z)$;
\item $R = \nu_{\rm obs}/\delta \nu$: the frequency resolving power, which sets the frequency resolution $\delta\nu$ and hence our capability of performing tomography;
\item $\sigma_{\rm pix}$: the noise variance per pixel (in Jy$\,\rm s^{1/2}/sr$).
\end{itemize}
We define three possible scenarios: the first, based on a CCAT-prime~\cite{Choi:2019rrt}-like survey, describes the capabilities of a [CII] detector of the current generation. Looking into the future, the most optimistic detector we consider is inspired by an AtLAST~\cite{Klaassen:2019,Booth:2024jkg,Booth:2024oem}-like survey, with a very good detector sensitivity, a long observational time and a wide sky area observed. For these two scenarios, we adopt similar parameters as the ones in Ref.~\cite{Bernal:2020lkd}. Finally, we consider an intermediate survey, which we label as reduced-AtLAST: this is characterized by the same noise sensitivity of the AtLAST-like surveys, but observes a much smaller sky area in a shorter observational time. As in Ref.~\cite{Bernal:2020lkd}, for each survey we consider four redshift bins within an  observed redshift range of $3.5<z<8.1$ for CCAT-prime and $1<z<5$ for ATLAST.

The specifications for all the detectors are summarized in Tab.~\ref{table:LIM_specs}. They are used in the following sections to estimate the LIM instrumental white noise for a single dish experiment, given by
\begin{equation}\label{eq:LIM_noise}
    N_X = \sigma_N^2 V_{\mathrm{vox}} = \frac{\sigma_{\mathrm{pix}}^2 V_{\mathrm{vox}}}{t_{\mathrm{pix}} N_{\mathrm{pol}} N_{\mathrm{feeds}}},
\end{equation}
where $V_{\mathrm{vox}}\propto \delta\nu\theta_{\rm FWHM}^2$ is the volume of each of the voxels defined on the LIM map, inside which intensity fluctuations are measured ($X = [{\rm CII}]$ in our case). The size of each voxel is defined by the frequency resolution in the radial direction, and by the pixel having area equal to the angular resolution $\theta_{\rm FWHM}$ on the plane of the sky. We assume a uniform observation over the survey volume, hence the full observational time is given by $t_{\rm obs} = t_{\rm pix}N_{\rm pix}$, where $N_{\rm pix}\propto \Omega_{\rm sky}/\theta^2_{\rm FWHM}$ is the number of pixels, and $t_{\rm pix}$ is the time spent observing each of them. As  is evident in Tab.~\ref{table:LIM_specs}, the noise power spectrum $N_X$, estimated in ${\rm Mpc^3 Jy^2sr^{-2}}$ units, is smaller for reduced-AtLAST than for AtLAST. This is due to the smaller sky area observed, which allows us to observe each voxel for a longer time, i.e.,~$t_{\rm pix}$ in Eq.~\eqref{eq:LIM_noise} is larger for reduced-AtLAST, reducing the noise variance per sterradian.
Finally, in Eq.~\eqref{eq:LIM_noise}, $\sigma_N$ is
the survey sensitivity in terms of the noise variance per voxel, 
$N_{\mathrm{feeds}}$ is the number of detectors in each antenna (which we set to 1) and $N_{\mathrm{pol}}$ is the number of possible antenna polarizations (set to 1 throughout the analysis).
Other noise sources, e.g.,~red noise, can be neglected at this stage.

\begin{table*}[t]
\centering
\begin{tabular}{|c|c|c|c|c|c|c|c|c|c|}
\hline\hline
{Experiment} & {$\,\,\,\,\,\nu_\mathrm{obs}\,\,\,\,\,$} & {$\,\,\,\,\,\Delta \nu\,\,\,\,\,$} & {R} & {$\,\,\,\,\sigma_\mathrm{FWHM}\,\,\,\,$} & {$t_\mathrm{obs}$} &{$\,\,\,\,\,\,\Omega_\mathrm{sky}\,\,\,\,\,\,$} & 
{$\sigma_\mathrm{pix}$} & {$\sigma_N$} & {$N_\mathrm{X}$}  \\ 
{ } & {[GHz]} & {[GHz]} & { } & {[arcsec]} & {[h]} &{[deg$^2$]} & {$\,\,\,$[Jy sr$^{-1}\sqrt{\mathrm{s}}$]$\,\,\,$} & {$\,\,\,\,\,\,$[Jy sr$^{-1}$]$\,\,\,\,\,\,$} &{[Mpc$^3\, \mathrm{Jy}^2 \mathrm{sr}^{-2}$]} \\
\hline\hline
CCAT-prime &  410 & 40 & 100 & 12.7 & 4000 & 8 & $27.1 \times 10^4$ & $5.7 \times 10^4$ & $22.4 \times 10^9$ \\ 
{} &  350 & 40 & {} & 14.8 & {} & {} & $13.8 \times 10^4$ & $2.5 \times 10^4$ & $6.6 \times 10^9$ \\ 
{} &  280 & 40 & {} & 19.1 & {} & {} & $7.1 \times 10^4$ & $1.0 \times 10^4$ & $1.98\times 10^9$ \\ 
{} &  220 & 40 & {} & 24.2 & {} & {} & $5.4 \times 10^4$ & $0.6 \times 10^4$ & $1.1 \times 10^9$ \\ 
\hline
AtLAST &  760 & 317 & 1000 &  2.0 & {\bf 10000} & {\bf 7500} & $15.0 \times 10^3$ & $3.9 \times 10^5$ & $13.5 \times 10^8$ \\ 
{} &  543 & 158 & {} &  2.8 & {} & {} & $7.5 \times 10^3$ & $1.4 \times 10^5$ & $5.3 \times 10^8$ \\ 
{} &  422 & 95 & {} &  3.6 & {} & {} & $4.8 \times 10^3$ & $0.7 \times 10^5$ & $2.7 \times 10^8$ \\ 
{} &  345 & 63 & {} &  4.4 & {} & {} & $3.4 \times 10^3$ & $0.4 \times 10^5$ & $1.4 \times 10^8$ \\ 
\hline
reduced-AtLAST &  760 & 317 & 1000 &  2.0 & {\bf 3000} & {\bf 600} & $15.0 \times 10^3$ & $2.0 \times 10^5$ & $3.6 \times 10^8$ \\
\hline

\hline\hline
\end{tabular}
\caption{Specifications for the LIM surveys used in this work (based on Ref.~\cite{Bernal:2020lkd}), alongside the calculated variance per voxel and instrumental noise power spectrum (see Eq.~\eqref{eq:LIM_noise}). Except for $t_{\rm obs}$ and $\Omega_{\rm sky}$, 'reduced-AtLAST' presents the same setup as AtLAST; for this reason, for brevity, we show  its specs for a single redshift bin, centered at $z=1.5$.}
\label{table:LIM_specs}
\end{table*}

\subsection{CMB}
To perform the velocity reconstruction described in Sec.~\ref{sec:vel_tom}, the LIM surveys described in the previous section need to be cross correlated with the measurements of a CMB experiment. In our analysis, we consider CMB-S4~\cite{CMB-S4:2016ple}, with typical angular resolution $\theta_\mathrm{FWHM, CMB}=1.4'$ and temperature sensitivity $\Delta_T =1.4\mu$K. 
Instrumental noise for CMB measurements is calculated using
\begin{equation}\label{eq:cmb_noise}
    N_l = \Delta_T^2 \exp\left[\frac{l(l+1)\theta_\mathrm{FWHM}^2}{8 \, \mathrm{ln}(2)} \right].
\end{equation}

Our choice is analogous to the one in Ref.~\cite{Tishue:2024}, where the velocity field is reconstructed by cross correlating the CMB with a galaxy survey, and it allows us to compare the performance of LIM and galaxies in constraining $\sum m_\nu$. The interested reader can find in Ref.~\cite{Tishue:2024} further details on kSZ and ML velocity reconstructions, as well as the cosmological constraints and how they depend on the choice of the CMB experiment.

We mention here that we model the CMB power spectrum, including the instrumental noise, without including foregrounds treatment. Including those will amount to a more conservative approach, as shown in Ref.~\cite{Tishue:2024}.

\subsection{Galaxy surveys}
To compare our LIM results to the benchmark set by galaxy surveys, we run the forecast analysis on neutrino masses also in the case of a spectroscopic and a photometric galaxy survey. For the former, we rely on the Dark Energy Survey Instrument (DESI) 5-year forecasts as depicted in Refs.~\cite{DESI:2016fyo, DESI:2023dwi}, while for the latter we adopt the Vera Rubin Observatory 10-year forecast (VRO10,~\cite{LSSTScience:2009jmu}). 
The specs for the galaxy surveys are summarized in Tab.~\ref{table:gal_specs}; in the case of VRO10, we set its photo-$z$ error to be $\sigma_z = 0.03\,(1+z)$. We account for the limited observed volume $V$ indicated in Tab.~\ref{table:gal_specs} by introducing a top-hat window function in the $k$ modes used to estimate the galaxy power spectrum.

\begin{table}
\renewcommand{\arraystretch}{1.2}
\begin{tabular}{|c|c|c|c|c|}
\hline\hline
{Experiment} & {z} & {$b_g$   } & {$n_\mathrm{gals}\,[\mathrm{Mpc}^{-3}]$} & {$V$ [Gpc$^3$]} \\ 
\hline\hline
DESI & 0.2 & 1.49 & $2 \times 10^{-3}$ & 5.3 \\
& 0.6 &  2.32 & $2 \times 10^{-4}$ & 20.4 \\ 
& 1.0 &  2.73 & $2 \times 10^{-4}$ & 18.3 \\ 
& 1.4 &  1.59 & $3 \times 10^{-4}$ & 18.7 \\ 
& 1.9 &  2.73 & $3 \times 10^{-4}$ & 2.3 \\ 
\hline 
VRO10 & 0.2 &  1.05 & $5 \times 10^{-2}$ & 5.2 \\ 
& 0.7 &  1.37 & $2 \times 10^{-2}$ & 43.6 \\ 
& 1.3 &  1.79 & $6 \times 10^{-3}$ & 65.9 \\ 
& 1.9 &  2.22 & $1.5 \times 10^{-3}$ & 86.3 \\ 
& 2.6 &  2.74 & $3 \times 10^{-4}$ & 119.6 \\ 
\hline\hline
\end{tabular}
\caption{Specifications for the galaxy surveys used in the comparison in Sec.~\ref{sec:analyis}; DESI specs are from Ref.~\cite{DESI:2023dwi}, while for VRO10 we used the same specs as Ref.~\cite{Tishue:2024}.}
\label{table:gal_specs}
\end{table}

\section{Methodology}
\label{sec:methods}

In this section, we summarize the formalism used to model the LIM power spectrum (Sec.~\ref{sec:lim_theory}), the velocity reconstruction (Sec.~\ref{sec:vel_tom}) and the Fisher forecast (Sec.~\ref{sec:fisher}).

\subsection{Line-Intensity Mapping}
\label{sec:lim_theory}

The quantities usually measured on LIM maps (see e.g., Refs.~\cite{Kovetz:2017agg,Bernal:2019jdo,Bernal:2022jap} for review) are the specific intensity $I_\nu \,[{\rm Jy/sr}]$ or, for experiments observing lower frequencies, the brightness temperature $T\,[{\rm K}]$, where the two can be related by the Rayleigh-Jeans relation\begin{equation}
    T(z) = \frac{c^2 (1+z)^2}{2 k_B \nu^2} I_\nu(z),
\end{equation}
$c$ being the speed of light and $k_B$ the Boltzmann constant, $\nu$ the rest frame frequency and $z$ the emission redshift.
In this work, we will adopt the $I_\nu$ formalism. 

For a given line, the intensity depends on the underlying source population through the line-luminosity density per comoving volume $\rho_{L_X}$, via
\begin{equation}
    I_\nu(z) = \frac{c}{4\pi\nu H(z)}\rho_{L_X}(z) \,\, , \,\, 
\end{equation}
where $X=$ [CII] in our case, and $H(z)$ is the Hubble expansion rate at the emission.
For most of the observed spectral lines at low $z$, we can assume that they originate inside galaxies that reside in dark matter halos. 
Therefore, we can express the average line luminosity density $\langle\rho_{L_X}\rangle(z)$ using the halo mass function $(dn/dM)(M,z)$~\cite{Tinker:2008ff} and mass-luminosity relation $L_X(M,z)$ through the halo model~\cite{Cooray:2002dia}, as
\begin{equation}
    \langle \rho_{L_X} \rangle(z) = \int dM L_X(M,z) \frac{dn}{dM}(M,z) .
\end{equation}

The summary statistic we work with is the LIM power spectrum, which is given by the Fourier transform of the two-point correlation function of the intensity spatial fluctuations around the mean value observed in the map, $\langle \delta_X\delta_X\rangle$. Since the line emission originates in halos, these fluctuations represent a biased tracer of the matter density fluctuations, $\delta_{\rm LIM}\propto b_X(z)\delta_m$, where $b_X(z)$ is the line luminosity-weighted bias. At linear order, we can approximate its value as
\begin{equation}\label{eq:LIM_bias}
    b_X(z) = \frac{\int dM L_X(M,z) b_h(M,z)\, ({dn}/dM)(M,z)}{\int dM L_X(M,z) \,({dn}/{dM})(M,z)},
\end{equation}
where $b_h(M,z)$ is the halo bias~\cite{Tinker:2010my}.  

Since observations are made in redshift space, the LIM power spectrum is affected by redshift-space distortions~\cite{Challinor:2011bk}. These are caused by peculiar velocities, which distort the observed position along the line-of-sight with respect to real space. At this stage, we neglect the role of the `Fingers of God' effect for small scales~\cite{Jackson:1971sky} and GR effects, such as lensing or Doppler. We model the RSD using the factor $F_{\rm RSD}$, accounting for the Kaiser effect~\cite{Kaiser:1987qv} for large scales
\begin{equation}\label{eq:F_RSD}
    F_{\rm RSD}(k,z) = \left( 1 + \frac{f(k,z)}{b_X(z)}\mu^2 \right),
\end{equation}
where $f(k,z)$ is the scale-dependent growth rate described in Sec.~\ref{sec:analyis}, $k$ is the module of the Fourier mode and $\mu = k_r/k$ is the cosine of the angle between the mode vector $\boldsymbol{k}$ and its line-of-sight (LoS) component $k_r$.

Overall, we model the LIM power spectrum using the halo model~\cite{Cooray:2002dia}, in which
\begin{equation}\label{eq:pLIM}
    P(k,\mu,z) = P_{\textrm{clust}}(k,\mu,z) + P_{\textrm{shot}}(z),
\end{equation}
where $P_{\rm shot}(z)$ is the shot noise and $P_{\textrm{clust}}(k,\mu,z)$ is the clustering part. This is obtained from the sum of two main components, 
\begin{equation}\label{eq:pclust}
     P_{\textrm{clust}}(k,\mu,z) = F_{\rm RSD}^2(k,z)\left[ P_{2h}^{XY}(k,\mu,z) + P_{1h}^{XY}(k,\mu,z) \right],
\end{equation}
namely the two-halo and one-halo terms as a function of the scale $k$, direction compared to the LoS $\mu$ and redshift $z$. 
The former term describes the correlation between different halos, it traces the matter power spectrum and it accounts for contributions from all the effects previously introduced. The latter, instead, describes the correlation within a single halo. In the context of cross-correlation power spectra between two tracers $X$ and $Y$, the terms are given by
\begin{align}\label{eq:1halo}
    P_{1h}^{XY} &= \int dM \frac{dn}{dM} F_X(k,M,z) F_Y(k,M,z) \\ \nonumber
    P_{2h}^{XY} &= \left[ \int dM \frac{dn}{dM} b_X(M,z) F_X(k,M,z) \right] \times \\ \nonumber 
    &\times\left[ \int dM \frac{dn}{dM} b_Y(M,z) F_Y(k,M,z) \right] P_{\mathrm{m}}(k,z) \\ \label{eq:2halo}
\end{align}
where $b_{X(Y)}$ is the line luminosity-weighted linear bias for the tracer
$X(Y)$ 
, $F_{X(Y)}$ is a profile function that depends on the tracer and $P_{\mathrm{m}}$ is the linear matter power spectrum. For line emission, the profile function is 
\begin{equation}\label{eq:line_profile}
    F_{\mathrm{X}}(M,z) = L_{\mathrm{X}}(M) u(k,M,z),
\end{equation}
where $u$ is the Fourier transform of the density profile of a halo of mass $M$. Specifically for  [CII], we can use the luminosity-weighted bias in Eq.~\eqref{eq:LIM_bias} and the profile
   $F_{\mathrm{CII}}(M,z) = \frac{c}{4\pi \nu H(z)} L_{\mathrm{CII}}(M,z) u_{\mathrm{NFW}}(k,M,z)$,
where $L_{\mathrm{CII}}(M,z)$ is the [CII] luminosity-halo mass relation~\cite{Silva:2014ira} and $u_{\rm NFW}$ is the Fourier transform of the Navarro-Frenk-White (NFW) density profile~\cite{Navarro:1996gj}. 

Since the one-halo term in Eq.~\eqref{eq:pclust} describes the non-linear contribution to the matter power spectrum, it turns out to be mainly relevant on
small scales. In the case of large scales, instead,
we can neglect this term and write the simplified expression 
\begin{equation}\label{eq:LIM PS 2halo}
    P_{\textrm{clust}}(k,\mu,z) = I_\mathrm{X}^2(z) b_X^2(z) F_{\rm RSD}^2(k,\mu,z)P_m(k,z).
\end{equation}
For convenience in the analysis in Sec.~\ref{sec:analyis}, we collect the degenerate parameters, by defining
\begin{equation}
    B_\mathrm{X}(z) \equiv I_\mathrm{X}(z) b_X(z),
\end{equation}
which accounts for both the line intensity and the line luminosity-weighted bias.

Finally, we can model the shot noise in Eq.~\eqref{eq:pLIM}, which stems from the fact that the line emissions trace the underlying dark-matter field in a discrete way, as
\begin{equation}\label{eq:shot}
    P_{\textrm{shot}}(z) = \left(\frac{c}{4\pi\nu H(z)}\right)^2 \int dM L_X^2(M,z) \frac{dn}{dM}(M,z).
\end{equation}
Note that, when this formalism is used to compute the cross correlation between different tracers $X\neq Y$, the shot noise term cancels out.

The power spectrum in Eq.~\eqref{eq:pLIM} represents the intrinsic signal one aims to observe. 
In practice, LIM experiments have limited angular resolution and probe a finite volume; these properties limit the smallest and largest accessible scales, respectively.
We model this fact by applying a window function to the power spectrum, given by
\begin{eqnarray}\label{eq:LIM_PS_full}
    \Tilde{P}(k,\mu,z) &=& W(k,\mu,z)P(k,\mu,z) =  \\ \nonumber &=& W_{\textrm{vol}}(k,\mu,z)W_{\textrm{res}}(k,\mu,z)P(k,\mu,z)
\end{eqnarray}
where 
\begin{align}
 W_{\textrm{vol}} =& \left( 1-\mathrm{exp} \left[ -\left( \frac{k}{k_\perp^\mathrm{min}}\right)(1-\mu^2)\right] \right) \times \\ \nonumber
 &\times \left( 1-\mathrm{exp} \left[ -\left( \frac{k}{k_\parallel^\mathrm{min}}\right)\mu^2\right] \right)
\end{align}
is the survey-area and 
\begin{equation}
W_{\textrm{res}}  = \mathrm{exp} \left(-k^2 \left[\sigma_\perp^2(1-\mu^2)+\sigma_\parallel^2\mu^2\right]\right)
\end{equation}
the instrument-response window functions in Fourier space, respectively. This, finally, sums to the noise power spectrum defined in Eq.~\eqref{eq:LIM_noise}. For details, see Ref.~\cite{Bernal:2022jap}.

\subsection{Velocity tomography}
\label{sec:vel_tom}

\begin{figure} [t]
	\centering
	\includegraphics[width=\columnwidth]{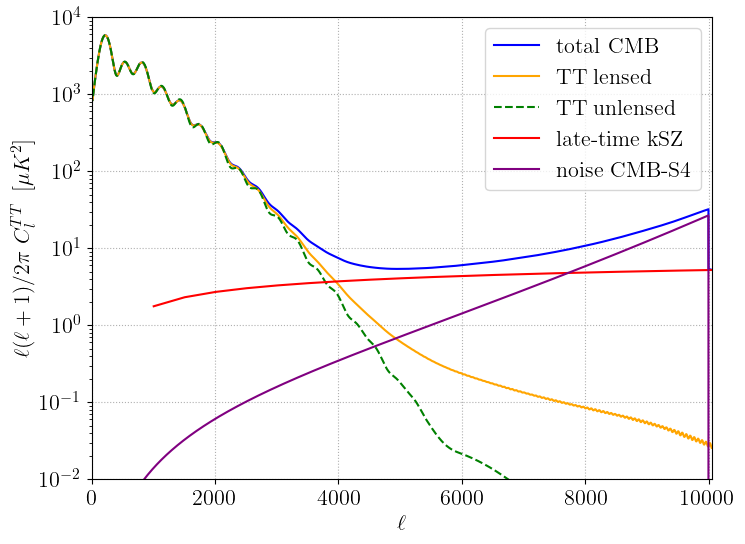}\\
	\caption{CMB power spectrum from primary CMB, gravitational lensing, late-time kSZ, and noise for CMB-S4.
 }
\label{fig:kSZ_PS}
\end{figure}

Velocity tomography leverages the cross-correlation between the CMB and a LSS tracer, such as galaxies or LIM, to probe the Universe on large scales. In this Section, we use the late-time kSZ\footnote{Patchy kSZ from reionization could also be analysed~\cite{McQuinn:2005ce, Park:2013mv}.} and ML effects, both related to secondary anisotropies in the CMB, relevant at $\ell\geq 4000$, to reconstruct the velocity field in the radial and transverse directions, respectively. As  will become clear throughout this Section, kSZ represents the dominant contribution; Fig.~\ref{fig:kSZ_PS} compares its effect with the CMB primary anisotropies, the CMB lensing and the forecasted noise for CMB-S4~\cite{CMB-S4:2016ple}.

To model the velocity signal, we use the linear order relation between the velocity and the matter density fields, based on the continuity equation
\begin{equation}\label{eq:continuity}
    \bm{v}(\bm{k},z) = \bm{\hat{k}}\frac{f(k,z)a(z)H(z)}{k}\delta(\bm{k},z)
\end{equation}
where $a(z)$ is the scale factor, and the other quantities were previously defined. By  means of the continuity equation, velocity can be considered as an unbiased tracer of the underlying matter density, whose potential sources the velocity field itself; this relation holds on large scales. Moreover, the dependence on the scale-dependent growth rate $f(k,z)$ in Eq.~\eqref{eq:continuity} makes velocity sensitive to the contribution of massive neutrinos (see more details in Sec.~\ref{sec:analyis}), and hence can be used as a complementary probe of the sum of their masses, $\sum m_\nu$.

In this context, $\boldsymbol{v}(\boldsymbol{k},z)$ can be cross correlated with the galaxy number density fluctuations $\delta_g$ or the LIM intensity fluctuations $\delta_{\rm CII}$, in a multi-tracer analysis. To reconstruct the 3D velocity field, we can express it in terms of its components along the LoS and in the transverse direction, as
\begin{equation}
    v = \mu^{-1} v_\parallel,\quad v = (\sqrt{1-\mu})^{-1}v_\perp.
\end{equation}
This leads us to express the velocity auto-power spectrum 
\begin{equation}\label{eq:P_vv}
    P_{vv}(k,z) = \left(\frac{b_v f(k,z)a(z)H(z)}{k} \right)^2 P_{\mathrm{m}}(k,z)
\end{equation}
where $b_v$ is the velocity reconstruction bias, which can be separated into the radial bias $b_\parallel=b_v\mu$ and the transverse bias $b_\perp =b_v\sqrt{1-\mu^2}$. As we detail in the following, $b_\parallel$ and $b_\perp$ arise from uncertainties in the theoretical modelling of the kSZ and ML effects, respectively. 

Similarly, we can express the cross power spectrum between the reconstructed velocity and a LSS tracer; for LIM, we can combine Eqs.~\eqref{eq:line_profile}, \eqref{eq:LIM_PS_full}, \eqref{eq:continuity} and write
\begin{equation}\label{eq:P_Xv}
\begin{aligned}
    P_{\mathrm{X}v} = &\left(\frac{b_v f(k,z)a(z)H(z)}{k} \right) \times \\
&  \times  B_\mathrm{X}(z) 
F_\mathrm{RSD} \sqrt{W(k,\mu,z)} P_{\mathrm{m}}(k,z).
\end{aligned}
\end{equation}

In our analysis in Sec.~\ref{sec:analyis}, we account for both the radial and transverse velocity fields and their cross-correlations; in the next subsections, we derive explicit expressions for each of them.

\subsubsection{kSZ effect}
\label{sec:kSZ_tom}

To model the kSZ effect~\cite{Zeldovich:1969ff,Sunyaev:1972eq,Birkinshaw:1984}, we follow Ref.~\cite{Smith:2018bpn}. 
The temperature anisotropy induced by kSZ is given by the LoS integral:
\begin{equation}\label{eq:TkSZ}
    T_{\mathrm{kSZ}} (\bm{\theta}) = K_* \int_0^{L} dr\, q_r(\chi_* \bm{\theta} + r \boldsymbol{\hat{r}}),
\end{equation}
where $\boldsymbol{\hat{r}}$ is the unit vector in the radial direction, $\chi_*$ the comoving distance associated with $z_*$, $\bm{\theta}$ is the angular direction on the sky and $q_r(\boldsymbol{x}) = \delta_e(\boldsymbol{x}) v_r(\boldsymbol{x})$ is the electron momentum field projected onto the radial direction. 
We define $K_* =  K(z=z_*)$ as the radial weight function in units of $[\mu$K\,Mpc$^{-1}]$ evaluated at $z_*$, where
\begin{equation}
    K(z) = -T_{\mathrm{CMB}} \sigma_T n_{e,0} x_e(z) e^{-\tau(z)} (1+z)^2.
\end{equation}
In the previous expression, $T_{\mathrm{CMB}}$ is the mean CMB temperature, $\sigma_T$ is the Thompson cross-section, $n_{e,0} = 0.19 \, \mathrm{m}^3$ is the mean free electron number-density today and $\tau$ is the optical depth from the observer to a scatterer with peculiar velocity $v$ located at a comoving distance $\chi$ at redshift $z$. 

As we motivate in Appendix~\ref{sec:kSZ_bispectrum}, to study the kSZ effect we rely on the squeezed bispectrum $\langle \delta_X(\boldsymbol{k})\delta_X(\boldsymbol{k'}) T(\boldsymbol{\ell})\rangle$, where $\delta_X(\boldsymbol{k})$ characterizes a large scale perturbation, while $T(\boldsymbol{\ell})$ and $\delta_X(\boldsymbol{k'})$ describe small scale modes. Since $T(\boldsymbol{\ell})$ is the Fourier transform of Eq.~\eqref{eq:TkSZ} and $q_r=\delta_e v_r$, the tree-level bispectrum can be rewritten by  means of the Wick theorem (see details in Ref.~\cite{Smith:2018bpn}), in terms of the product between the $P_{Xe},\,P_{Xv}$ cross-power spectra.
This introduces the so-called ``kSZ optical depth degeneracy": the cross-power spectrum between the LSS tracer $X$ and, respectively, the velocity field and the electron density perturbation, can in principle be estimated independently; however, a multiplicative factor can be exchanged between the two while keeping the signal the same, introducing a degeneracy.
This can be encapsulated into the radial-velocity reconstruction bias~\cite{Hotinli:2023ywh}
\begin{equation}\label{eq:b_par}
\begin{aligned}
    b_\parallel =& \int dk_S k_S \,P^{\rm real}_{Xe}(k_S)\frac{ P_{Xe}^{\rm fid}(k_S)}{P_{XX}^{\rm obs}(k_S)C_{\ell=k_S\chi}^{\rm obs}}\,\times \\
    &\times \left[\int dk_S k_S\,P_{Xe}^{\rm fid}(k_S)\frac{P^{\rm fid}_{Xe}(k_S)}{P_{XX}^{\rm obs}(k_S)C_{\ell=k_S\chi}^{\rm obs}}\right]^{-1},
\end{aligned}
\end{equation}
where $C_{\ell}^{\mathrm{obs}}$ is the total CMB TT power spectrum, including the instrumental noise, and $P_{XX}^{\mathrm{obs}}$ is the observed auto-power spectrum of the LSS tracer $X$, including all relevant sources of noise. For LIM, $P_{XX}^{\mathrm{obs}}$ includes the shot noise term in Eq.~\eqref{eq:shot} and the instrumental noise in Eq.~\eqref{eq:LIM_noise}. 

Finally, $P_{Xe}^{\rm real}$ is the power spectrum intrinsic to the observations, while $P_{Xe}^{\rm fid}$ represents the fiducial model adopted in the analysis. In this sense, $b_\parallel$ accounts for uncertainties in the modeling of the electron distribution on small scales. To model $P_{Xe}$, we use Eqs.~\eqref{eq:1halo} and~\eqref{eq:2halo}, where we set
\begin{equation}\label{eq:density_profile_electron}
    F_e(M,z)  = \frac{M}{\overline{\rho}_m} u_{\mathrm{gas}}(k,M,z),
\end{equation}
where $\overline{\rho}_m$ is the present-day mean matter density and $u_{\rm gas}$ is the Fourier transform of the electron density profile. This is modeled from Ref.~\cite{Battaglia:2016xbi}; compared with the NFW profile, the density profile lacks a central cusp and shows a steeper fall-off at large radii. To estimate $P_{{\rm X}e}$ based on Eq.~\eqref{eq:2halo}, we finally approximate the electron bias to be equal to the halo bias.

\begin{figure}[ht!]
	\centering	\includegraphics[width=\columnwidth]{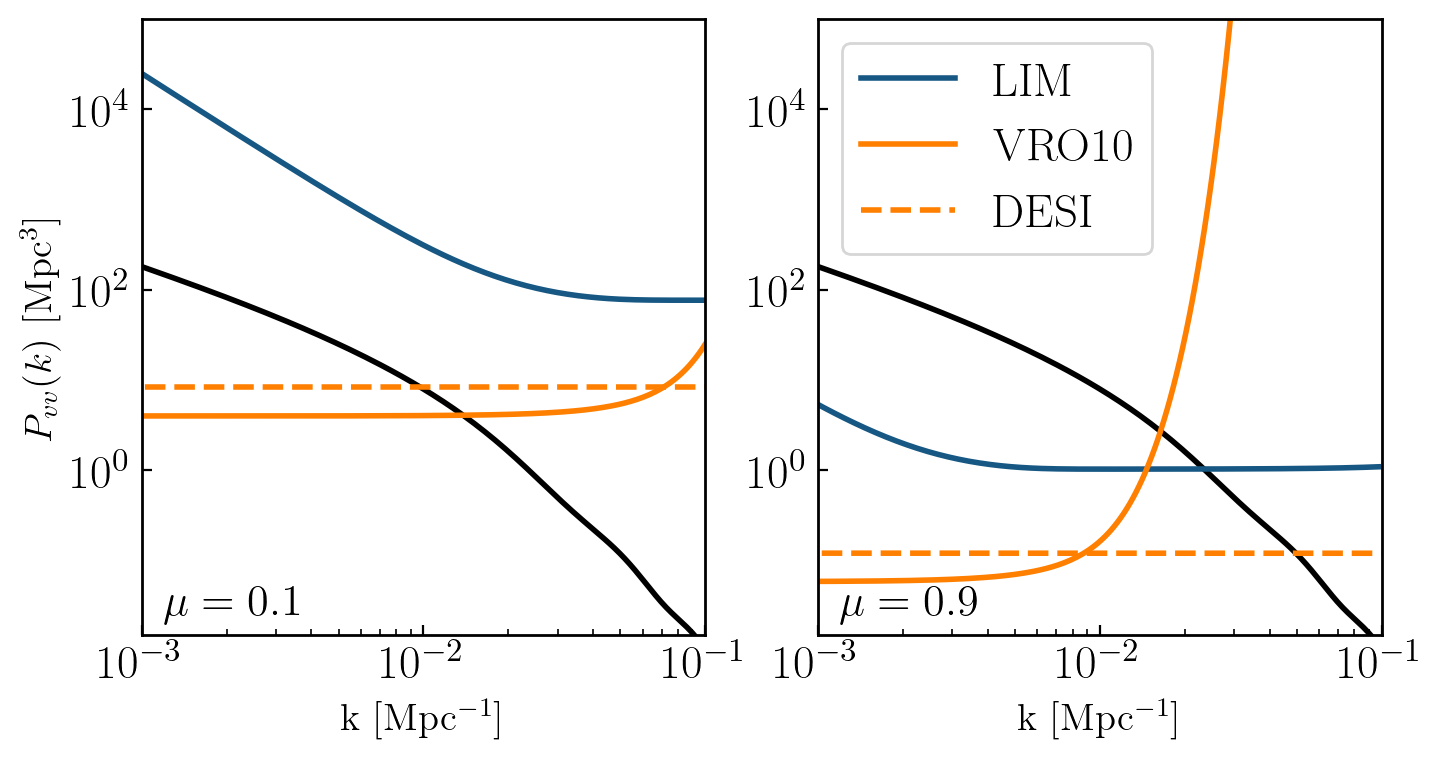}\\
	\caption{Velocity auto-power spectrum (in black) compared with kSZ reconstruction noise for LIM (blue, see specs in Sec. \ref{sec:experiments}) and galaxy surveys (VRO10 and DESI, orange). We use the bin centered at $z=1.5$ for reduced-AtLAST, $z=1.3$ for VRO10 and $z=1.4$ for DESI. Two cases are shown, in the nearly-transverse ($\mu=0.1$) and nearly-radial ($\mu=0.9$) directions respectively.}
\label{fig:ksz_noise}
\end{figure}

Following the formalism in Ref.~\cite{Smith:2018bpn}, the bispectrum can be used to reconstruct the long-wavelength radial velocity modes. This is done by marginalizing over the small scale modes of the CMB temperature and tracer density  fluctuations; therefore, the noise power spectrum of the reconstructed radial velocity is derived to be~\cite{Smith:2018bpn}
\begin{equation}\label{eq:noise_par}
    N_{v_\parallel}(k_{L_\parallel}) = \frac{\chi_*^2}{K_*^2} \left[ \int_{k_S^{\rm min}}^{k_S^{\rm max}} \frac{dk_S k_S}{2\pi} \frac{P_{\mathrm{X}e}^2(k_S, k_{L_\parallel})}{P_{\mathrm{XX}}^{\mathrm{obs}}(k_S, k_{L_\parallel}) C_{\ell=k_S \chi_*}^{\mathrm{obs}}} \right]^{-1},
\end{equation}
where, in the squeezed limit, we used $k_{Lr} = k_{Sr}$. {In our analysis in Sec.~\ref{sec:analyis} we adopt logspaced $k_S$ values between $k_S^{\rm min} = 0.1\,{\rm Mpc}^{-1}$ and $k_S^{\rm max} = 5\,{\rm Mpc}^{-1}$.\footnote{The upper limit is chosen conservatively, since including smaller scales would lead to smaller noise, particularly at lower redshifts.} }

Now, using the relation $v(\boldsymbol{k}_L) = \mu^{-1} v_r(\boldsymbol{k}_L)$, we can estimate the noise for the 3D velocity field,
\begin{equation}\label{eq:Nvv_ksz}
    N_{vv}(k_L, \mu) = \mu^{-2} N_{v_\parallel}(k_L, \mu).
\end{equation}

We show in Fig.~\ref{fig:ksz_noise} the reconstruction noise from kSZ tomography, both in the case of LIM (reduced-AtLAST, see Sec.~\ref{sec:experiments}) and a photometric and spectroscopic galaxy survey (VRO10 and DESI, see Sec.~\ref{sec:experiments}). The comparison with VRO10 reveals that LIM is advantageous for kSZ at small scales: while for photometric galaxy surveys $N_{vv}$ increases significantly due to photometric redshift errors, in the LIM case it saturates due to the fine frequency resolution, which translates into fine redshift resolution. 
This is similar to what happens when relying on a spectroscopic galaxy survey, e.g.,~DESI.
On the other hand, in the transverse direction ($\mu \sim 0$), LIM is limited compared to galaxy surveys by its observed volume (on large scales) and its angular resolution (on small scales). 

\subsubsection{ML effect}
\label{sec:ML_tom}

When galaxy clusters, or other potential-well sources (which we refer to as {\it lenses}), move in the plane orthogonal to the LoS, they can source peculiar velocities in the transverse direction. This produces in the CMB a typical dipole pattern, which is referred to as the  ML effect~\cite{Birkinshaw:1983,Tuluie:1995pg,
Aghanim:1998ux,Cooray:2002ee}; estimators for such quantity have been built in the context of CMB lensing, see e.g.,~Ref.~\cite{Hu:2001kj}, and applied in detail to the ML case in Ref.~\cite{Cayuso:2021ljq}.

ML provides CMB temperature fluctuations that can be modeled as~\cite{Hotinli:2018yyc}
\begin{equation}\label{eq:TML}
    T_{\mathrm{ML}} (\bm{\theta}) = -2 \int_0^L dr\, \dot{\Phi}(\bm{x}) =  -2 \int_0^L dr \bm{\nabla}_\perp \Phi(\bm{x}) \cdot \bm{v}_\perp(\bm{x}),
\end{equation}
where $\bm{x} = \chi_* \bm{\theta} + r \mathbf{\hat{r}}$ and $\bm{v}_\perp$ the peculiar comoving velocity in the transverse direction. Since we focus on the large scales, we can assume this to be linear and curl-free, and express it as 
\begin{equation}
    \bm{v}_\perp(\bm{x}) = \bm{\nabla}_\perp \Upsilon(\bm{x}),
\end{equation}
$\Upsilon$ being the large scale potential that sources it.
Finally, in Eq.~\eqref{eq:TML}, CMB photons are being deviated by the local Weyl non-linear gravitational potential (the one that evolves in time), modeled in synchronous gauge as $\Phi({\boldsymbol{r}})\sim 3\Omega_mH_0^2\delta/[a^2k^2]$,  where $\Omega_m$ is the matter density parameter, $H_0$ is the Hubble parameter and $a = (1+z)^{-1}$ is the scale factor. 

Similar to the case of kSZ described in the previous Section, we can use ML tomography to reconstruct the velocity field, this time the power of this reconstruction being in the transverse direction. In this case, the potential $\Phi(\boldsymbol{x})$ represents the small-scale quantity that gets marginalized in the process of velocity reconstruction, which in turn is related to the large-scale potential $\Upsilon$ that sources the velocity field. The noise in the reconstruction, in the squeezed limit $\bm{k} \ll \bm{k}_S, \bm{\ell}/\chi_{*}$, is hence given by
\begin{equation}
    N_{\Upsilon\Upsilon}(k_{L_\perp}) = \frac{\chi_*^2}{2 k_{L_\perp}^2} \left[ \int_{k_S^{\rm min}}^{k_S^{\rm max}} \frac{dk_S k_S^3}{2\pi} \frac{P_{\mathrm{\Phi\Phi}}^2(k_S)}{P_{\mathrm{\Phi\Phi}}^{\mathrm{obs}}(k_S) C_{\ell=k_S \chi_*}^{\mathrm{obs}}} \right]^{-1},
\end{equation}
where $k_{L_\perp} = k_L \sqrt{1-\mu^2}$, $P_{\mathrm{\Phi\Phi}}$ is the gravitational potential power spectrum and $P_{\mathrm{\Phi\Phi}}^{\mathrm{obs}}(k,\mu) = P_{\mathrm{\Phi\Phi}}(k) + N_{\mathrm{\Phi\Phi}}(k,\mu)$.
In the context of the LIM analysis performed in Sec.~\ref{sec:analyis}, we express it as\footnote{The full derivation of the gravitational potential noise can be found in Appendix~\ref{sec:N_phiphi}.} 
\begin{equation}\label{eq:Nphiphi}
    N_{\Phi\Phi} = \left[ \frac{(1+z) \rho_{m,0}}{2} \right]^2 \frac{1}{k^4  B_{{X}}^2} (P_{\mathrm{shot}}W(k,\mu) + N_X).
\end{equation}
The LIM shot noise $P_{\rm shot}$, window function $W(k,\mu)=W_{\rm res}(k,\mu)W_{\rm vol}(k,\mu)$ and instrumental noise $N_X$ are defined in Sec.~\ref{sec:lim_theory}. 
\begin{figure} [ht!]
	\centering
	\includegraphics[width=\columnwidth]{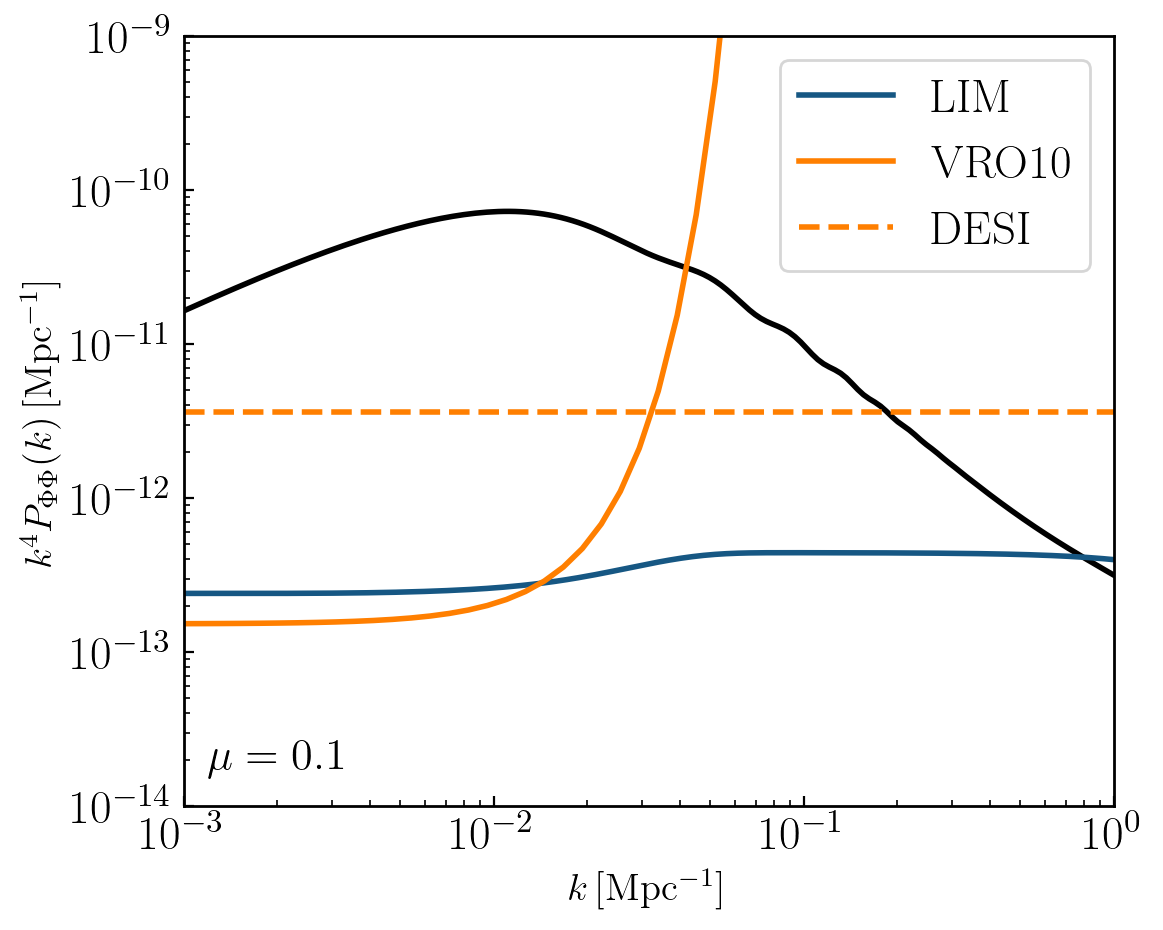}\\
	\caption{Gravitational potential power spectrum and noise (both multiplied by $k^4$ for aesthetic purposes) for LIM (blue) and compared with galaxy surveys (orange); the redshift bins are the same as in Fig.~\ref{fig:ksz_noise}. 
 See Sec.~\ref{sec:experiments} for detector properties.
 }
\label{fig:Nphiphi}
\end{figure}
Moreover, we account for uncertainties in the model of the small-scale gravitational potential in the bias factor $b_\perp$, which, analogously to $b_\parallel$ in Eq.~\eqref{eq:b_par}, is defined as~\cite{Hotinli:2021hih}
\begin{equation}\label{eq:b_perp}
\begin{aligned}
    b_\perp =& \int \frac{dk_S k_S \, P_{X\Phi}^{\rm fid}(k_S)P^{\rm real}_{X\Phi}(k_S)}{P_{XX}^{\rm obs}(k_S)C_{\ell=k_S\chi}^{\rm TT,tot}}\times \\
    &\times \left[\int dk_S k_S\frac{P_{X\Phi}^{\rm fid}(k_S)P^{\rm fid}_{X\Phi}(k_S)}{P_{XX}^{\rm obs}(k_S)C_{\ell=k_S\chi}^{\rm TT,tot}}\right]^{-1}.
\end{aligned}
\end{equation}

For the generality of this discussion, we note that in the case of galaxy surveys, the $N_{\Phi\Phi}$ expression in Eq.~\eqref{eq:Nphiphi} becomes
\begin{equation}
    N_{\Phi\Phi} = \left[ \frac{(1+z) \rho_{m,0}}{2} \right]^2 \frac{1}{k^4 b_g^2} \,\frac{1}{n_g W^2(k,\mu)},
\end{equation}
in which the last factor represents the galaxy shot noise.\\
We show the reconstructed $k^4N_{\Phi\Phi}$ from LIM and galaxy surveys in the transverse direction in Fig.~\ref{fig:Nphiphi}. Already from this plot, we can highlight some peculiar features LIM presents, that characterizes its different capability in the reconstruction of the potential, and hence of the transverse velocity via the ML effect. First of all, we note the different behaviour the noise inherits from the window function: when utilizing the 3D-$k$ value, the fine frequency resolution of LIM makes its behaviour similar to the one of the spectroscopic survey, with a slight decrement on scales $k\gtrsim 10^{-2}\,{\rm Mpc^{-1}}$. This can be traced to the angular resolution of the experiment that we are considering (see Tab.~\ref{table:LIM_specs}), which is not as fine as the galaxy one. Moreover, since LIM shot noise is estimated from the integrated luminosity function (see Eq.~\eqref{eq:shot}), up to the faintest sources that are not captured in galaxy surveys, its overall amplitude is smaller than the DESI case.

\begin{figure} [t]
	\centering
	\includegraphics[width=\columnwidth]{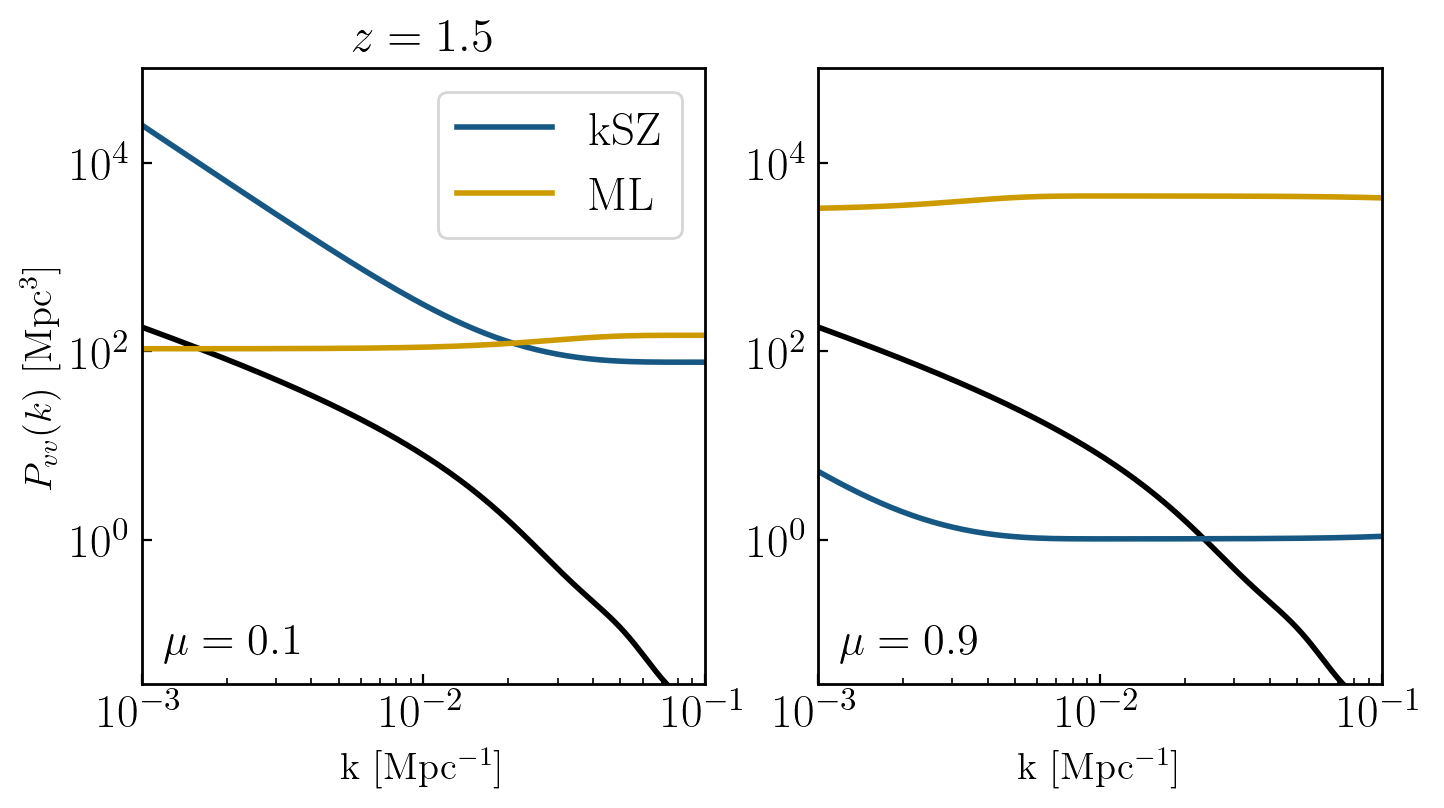}\\
	\caption{Velocity power spectrum (black) and reconstruction noise using kSZ effect (blue) and ML effect (yellow) on the full velocity field, in the nearly-radial and nearly-transverse directions. The plots are produced for the redshift bin $z=1.5$ of the reduced-AtLAST survey (see Tab.~\ref{table:LIM_specs}), and with CMB-S4 specifications for the noise spectra.
 }
\label{fig:Pvv_Nvv}
\end{figure}

Finally, we convert the noise in the reconstructed potential to noise in the reconstructed velocity field, using
\begin{equation}\label{eq:noise_perp}
    N_{v_\perp}(k_{L_\perp}) = k_L^2 N_{\Upsilon\Upsilon}(k_{L_\perp}).
\end{equation}
Similarly to the kSZ effect in Eq.~\eqref{eq:Nvv_ksz}, we can relate this to the noise for the total velocity field
\begin{equation}
    N_{vv}(k_L, \mu) = (\sqrt{1-\mu^2})^{-2} N_{v_\perp}(k_L, \mu).
\end{equation}
We show the reconstructed velocity noise in Fig. \ref{fig:Pvv_Nvv}, both using the kSZ effect and the ML effect. As expected, the kSZ effect enables a better reconstruction in the radial direction, while the ML effect can potentially better probe the transverse velocity. However, the ML noise level is overall larger than the one achievable with kSZ.
Figs.~\ref{fig:Nphiphi} and~\ref{fig:Pvv_Nvv} can be compared with Fig.~2 in Ref.~\cite{Hotinli:2021hih}, where the case of galaxy surveys is shown. 

\subsection{Fisher analysis}
\label{sec:fisher}

Based on previous studies in Refs.~\cite{Munchmeyer:2018eey,Hotinli:2021hih,Sato-Polito:2020cil}, we forecast the constraining power of the density and velocity field on cosmological parameters using the Fisher forecast formalism.
For a set of parameters ${\theta_1,...\theta_N}$, we build our Fisher matrix as
\begin{equation}\label{eq:fisher}
    F_{ij} = \frac{V}{2} \int_{k_L^{\rm min}}^{k_L^{\rm max}} \int_{-1}^{1} \frac{k^2 dk d\mu}{4\pi^2} \mathrm{Tr} \left[ C^{-1} \frac{dC}{d\theta_i} C^{-1} \frac{dC}{d\theta_j} \right],
\end{equation}
where $V$ is the survey volume, which for a LIM survey is $V = \Omega_{\rm sky}\chi^2(z)(\chi_{\rm max}-\chi_{\rm min})$ ($\chi_{\rm max,min}$ being the comoving distances associated with the boundaries of the redshift bin, centered in $z$). 
{We set $k_L^{\rm min} = 3 \cdot 10^{-4}\,{\rm Mpc}^{-1}$; however, access on the largest scales is limited by the window function $W_\mathrm{vol}$ (discussed in Sec.~\ref{sec:lim_theory}), due to the finite survey volume.}
In the following, we run the analysis using logspaced $k_L$ values between $k_L^{\rm min}$ and $k_L^{\rm max} = 0.1\,{\rm Mpc}^{-1}$, so that the $k_L$ range is complementary to $k_S$ used to estimate the noise in Sec.~\ref{sec:vel_tom}. $C=S+N$ is the total covariance matrix defined by the sum of the signal and noise covariances. 
These matrices contain the information coming from the LIM auto-power spectrum and from its use in reconstructing velocities in the LoS direction and the transverse plane. We build them using the wavenumber vector basis $\boldsymbol{k} = (k\mu, k\sqrt{1-\mu^2}, 0)$.\footnote{To overcome the degeneracy that exists between the two directions of the transverse plane, we arbitrarily choose the vector base in which the transverse velocity components are $\boldsymbol{v}_\perp = (v_\perp,0)$. The analysis is independent of this choice.}
In this base, the signal matrix $S$ is defined based on Eqs.~\eqref{eq:LIM PS 2halo},~\eqref{eq:P_vv} and~\eqref{eq:P_Xv}, as
\begin{equation}
    S_{ij}(k,\mu,z) = c_ic_j P_{m}(k,z),
\end{equation}
where the coefficients are given by $c_i = (c_\parallel, c_\perp, 0, c_X)$, namely
\begin{align}
    c_\parallel &\equiv b_\parallel \mu \frac{faH}{k}, \\
    c_\perp &\equiv b_\perp \sqrt{1-\mu^2} \frac{faH}{k}, \\
    c_X &\equiv [b_\mathrm{X}(z) + f\mu^2] I_\mathrm{X}(z) \sqrt{W(k,\mu,z)},
\end{align}
in which each block has size $N_k \times N_\mu$, and this is calculated for each redshift bin.
On the other side, the noise matrix $N$ is defined assuming that different measurements are uncorrelated, hence only diagonally
\begin{equation}
    N(k,\mu,z) = \mathrm{diag} \{N_\parallel, N_\perp, N_\perp, N_\mathrm{X}\},
\end{equation}
where the elements can be found in Eqs. \eqref{eq:noise_par}, \eqref{eq:noise_perp},~\eqref{eq:LIM_noise}. When the LIM auto-power spectrum is studied alone, the $S$ and $N$ matrices reduce to the lines and columns related with $c_X$ and $N_X$.

We run our LIM forecast analysis over the parameter set
$\theta_i \in \{H_0, \Omega_bh^2, \Omega_ch^2, A_s, n_s, \tau, \Sigma m_\nu, b_\parallel, b_\perp, B_\mathrm{X} \}$, which includes the standard $\Lambda$CDM parameters, the summed neutrino masses, the velocity bias parameters and the astrophysical parameters that define the line emission in Sec.~\ref{sec:lim_theory}.\footnote{Ref.~\cite{Tishue:2024} performs a similar analysis relying on galaxies instead of LIM. They adopt our same cosmological and velocity-bias parameters, but they further account for the bias parameters $\{b_g, b_2, b_{\rm RSD}\}$, where $b_g$ and $b_2$ are the linear and the second order term in the galaxy bias expansion. Moreover, $b_{\rm RSD}$ is introduced to account for uncertainties in the modeling of the RSD effect; this term multiplies $f\mu^2$ in $c_X$. While we fix this parameter throughout our analysis, the authors of Ref.~\cite{Tishue:2024} vary it to account for its effect.} We use the fiducial values for cosmological parameters: $\{67.5\,\mathrm{km/s/Mpc}, 0.022, 0.12, 2.2 \times 10^{-9}, 0.965, 0.06, 0.06\,\mathrm{eV}\}$, where for the neutrino masses we assume the normal hierarchy and take the fiducial value to be the minimal value allowed in this case. For velocity biases, we consider the fiducial to be unity, meaning that we rely on the small-scale fiducial models chosen in the analysis, see Eqs.~\eqref{eq:b_par},~\eqref{eq:b_perp}. 
Finally, the fiducial value of $B_\mathrm{X}=I_X(z)b_X(z)$ is estimated in each redshift bin of the LIM survey under consideration based on the underlying astrophysical model.

We tested that all the noise contributions vary negligibly when the parameter set is varied; therefore, the derivatives in Eq.~\eqref{eq:fisher} are numerically computed only with respect to the signal term. We estimate the Fisher matrices $F_z$ in each of the redshift bins of the surveys under analysis, defined in Sec~\ref{sec:experiments}. 
We then sum over the matrices to combine information from different redshift bins. To do so, $B_X(z)$ is treated as a set of four independent parameters, one for each bin.
Once we marginalize over other bias and cosmological parameters, this gives us the constraining power on neutrino masses in Sec.~\ref{sec:analyis}.

\section{Analysis and results} 
\label{sec:analyis}

In this section, we first of all summarize how massive neutrinos affect the different ingredients of our analysis (Sec~\ref{sec:massive_nu}), and then describe our results for the constraints on the sum of neutrino masses from the CII auto-power spectrum and velocity reconstruction (Sec~\ref{sec:nufromLIM}). We consider how this compares to galaxy surveys, and finally combine our LIM results with information coming from the CMB (Sec~\ref{sec:finalresults}). The entire code used in this work is available on Github\footnote{\url{https://github.com/shmuelig/LSS-Velocities}}.

\subsection{Effect of massive $\nu$ on the observables} \label{sec:massive_nu}

The presence of massive neutrinos has a significant impact on the overall matter density and matter power spectrum, see e.g.,~Ref.~\cite{Lesgourgues:2006nd} for review. At low $z$, massive neutrinos are non-relativistic, hence their contribution is accounted for in the $\Omega_m$ budget and their perturbations closely follow the dark matter ones. At high $z$, instead, they behave as radiation: on one side, this leads to a suppression in the growth of small scales due to their free streaming; on the other, their presence changes the time of matter-radiation equality. Moreover, while they affect the background evolution in the Friedmann equation $(\dot{a}/a)^2\propto (\bar{\rho}_{\rm CDM} + \bar{\rho}_{\rm b}+\bar{\rho}_{\rm \nu})$, at high $z$ they do not cluster and hence do not contribute to the Poisson equation that evolves $\delta\rho\ = (\bar{\rho}_{\rm CDM}+\bar{\rho}_{\rm b})\delta_{{\rm CDM}}$~\cite{Bond:1980ha}. This introduces further modifications in the evolution of the density perturbations, which is captured by defining a scale-dependent growth rate $f(k,a) = {d\,\mathrm{ln}\delta_{\rm CDM}(k,a)}/{d\,\mathrm{ln}a}$. 

\begin{figure*}[t]
    \centering
\includegraphics[width=2\columnwidth]{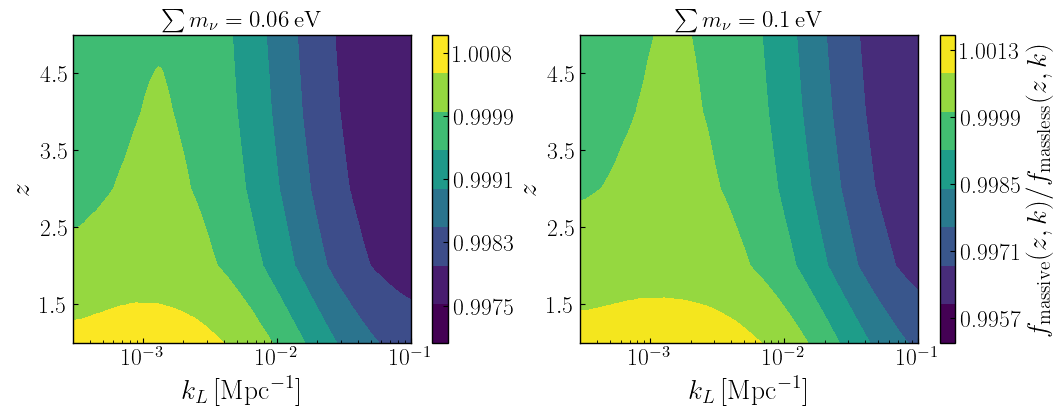}
    \caption{Ratio between the growth factor in the case of massive $\sum m_\nu = \{0.06,0.1\}$\,eV and massless neutrinos in the range of scales of interest for the analysis. The scale dependent effects are summarized in the main text, see also~Ref.~\cite{Lesgourgues:2006nd} for details.}
    \label{fig:growth}
\end{figure*}

In Fig.~\ref{fig:growth}, we summarize these effects by plotting the ratio between $f(k,z)$ in the massless and massive neutrino cases, obtained using the Boltzmann code \texttt{CLASS}~\cite{Lesgourgues:2011re}. The plot shows the cases of $\sum m_\nu= \{0.06,0.1\}\,{\rm eV}$, namely the limiting values in the cases of  normal and inverted hierarchies; the $(k,z)$ axes are limited to the values of interest in Sec.~\ref{sec:nufromLIM}. First of all, we note the suppression on small scales, which becomes stronger while increasing $\sum m_\nu$. The evolution of the density perturbations on these scales is affected by the neutrino free streaming before the non-relativistic transition. On the other hand, $f_{\rm massive}(z,k)/f_{\rm massless}(z,k)>1$ on large scales and high redshift: this is due to the fact that the effective value of $\Omega_m = \Omega_c + \Omega_b + \Omega_\nu$ increases with larger $\sum m_\nu$, leading to a faster growth. The position of the peak in Fig.~\ref{fig:growth} depends on the transition between the relativistic and non-relativistic regimes (compare with e.g.,~Fig.~13 in Ref.~\cite{Lesgourgues:2011re}).

The effects previously described impact both the LIM and velocity power spectra adopted in our analysis. The LIM power spectrum inherits the different shape of the matter power spectrum, as described in Eq.~\eqref{eq:pLIM}. The velocity power spectrum, beyond accounting for these effects, is also directly affected by the scale-dependent growth factor through the velocity bias in Eq.~\eqref{eq:P_vv}. 

\begin{table}
\renewcommand{\arraystretch}{2}
  \begin{tabular}{l|ccc}
  \hline\hline
    \toprule
    \multirow{1.2}{*}{Experiment} &
      \multicolumn{3}{c}{$\sigma_{\Sigma m_\nu}$ [meV]} \\
      & {$P_\mathrm{XX}$} & {$P_\mathrm{XX}+P_{v\mathrm{X}}+P_{vv}$} & {$\,\,$+$1\% \, b_\parallel$ prior} \\
      \midrule
    CCAT-prime & 7788 & - & - \\
    reduced-AtLAST & 166.6 & 162.8 & 148.9 \\
    AtLAST & 51.0 & 50.0 & 49.0 \\
    \hline
    DESI & 290.5 & 193.0 & 183.4 \\
    VRO10 & 170.3 & 125.0 & 122.6 \\
    \bottomrule
    \hline\hline
  \end{tabular}
  \caption{Forecasted marginalized 68\% confidence-level constraints on $\Sigma m_\nu$ for different experiments used in this work. The first column corresponds to forecasts from tracer $X$ only, in the second column we add velocity data and cross correlations with tracer $X$ and in the third we add the 1\% prior of $b_\parallel$ as discussed in the text.}
\label{table:sigma_mnu_summary}
\end{table}

\subsection{$\sum m_\nu$ constraints from LIM}\label{sec:nufromLIM}

\begin{figure} [t]
	\centering
	\includegraphics[width=\columnwidth]{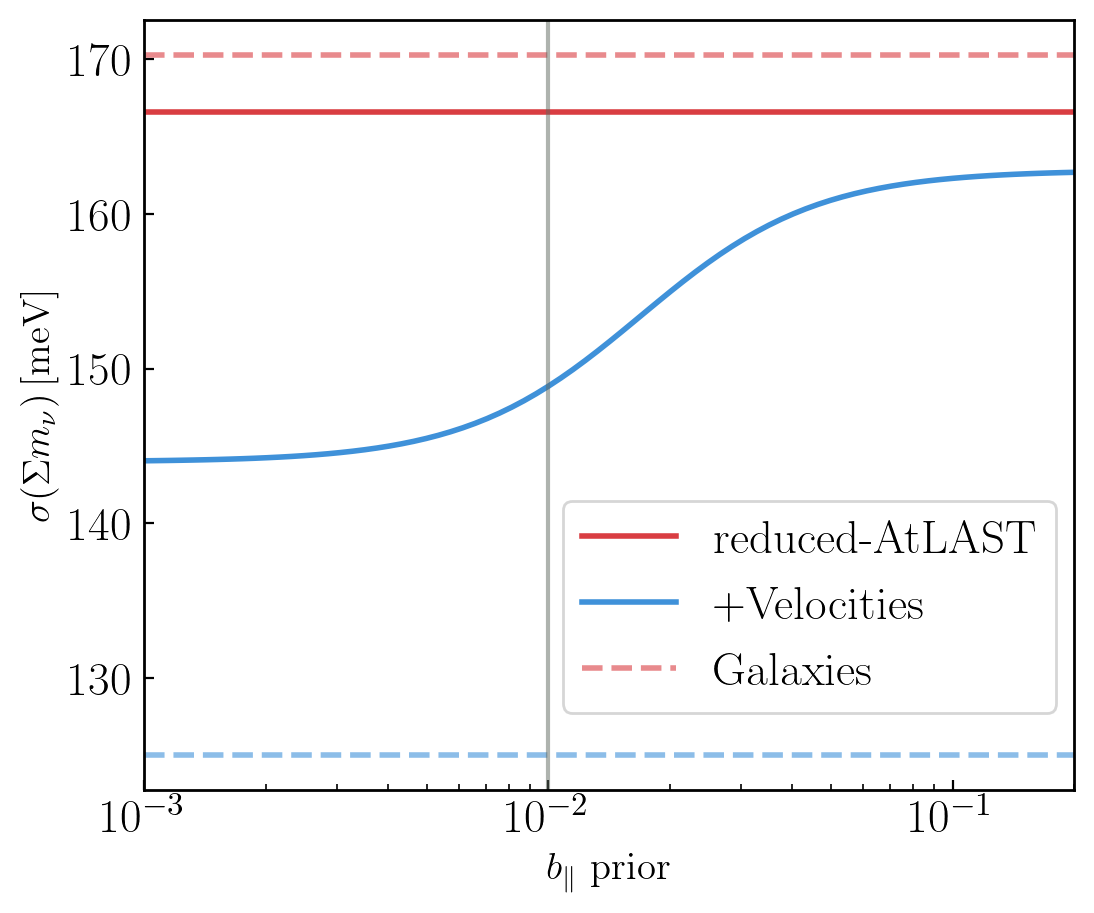}\\
	\caption{Constraints of the summed neutrino masses as function of the prior value set for $b_\parallel$. Solid lines show constraints from LIM survey reduced-AtLAST and adding velocity information, where matching colors dashed lines are from galaxy survey VRO10. The vertical line shows the reasonable prior according to Ref.~\cite{Madhavacheril:2019buy}.}
\label{fig:sigm_mnu_bpar}
\end{figure}

First of all, we run the Fisher forecast described in Sec.~\ref{sec:fisher} to constrain $\sum m_\nu$ from the CII auto-power spectrum; Tab.~\ref{table:sigma_mnu_summary} summarizes our results. While CCAT-prime can hardly constrain this parameter, providing $\sigma_{\Sigma m_\nu} \sim 8\,{\rm eV}$, reduced-AtLAST will reach 
a forecasted $\sigma_{\Sigma m_\nu} = 166\,{\rm meV}$. This value, as Fig.~\ref{fig:sigm_mnu_bpar} shows, is comparable with forecasts for galaxy surveys such as VRO10. Reaching the full AtLAST configuration will provide $\sigma_{\Sigma m_\nu} \sim 50\,{\rm meV}$, with a potential rejection of the inverted hierarchy at $>2\sigma$ driven by LIM maps alone. 

\begin{figure*} [ht!]
	\centering
	\includegraphics[width=2\columnwidth]{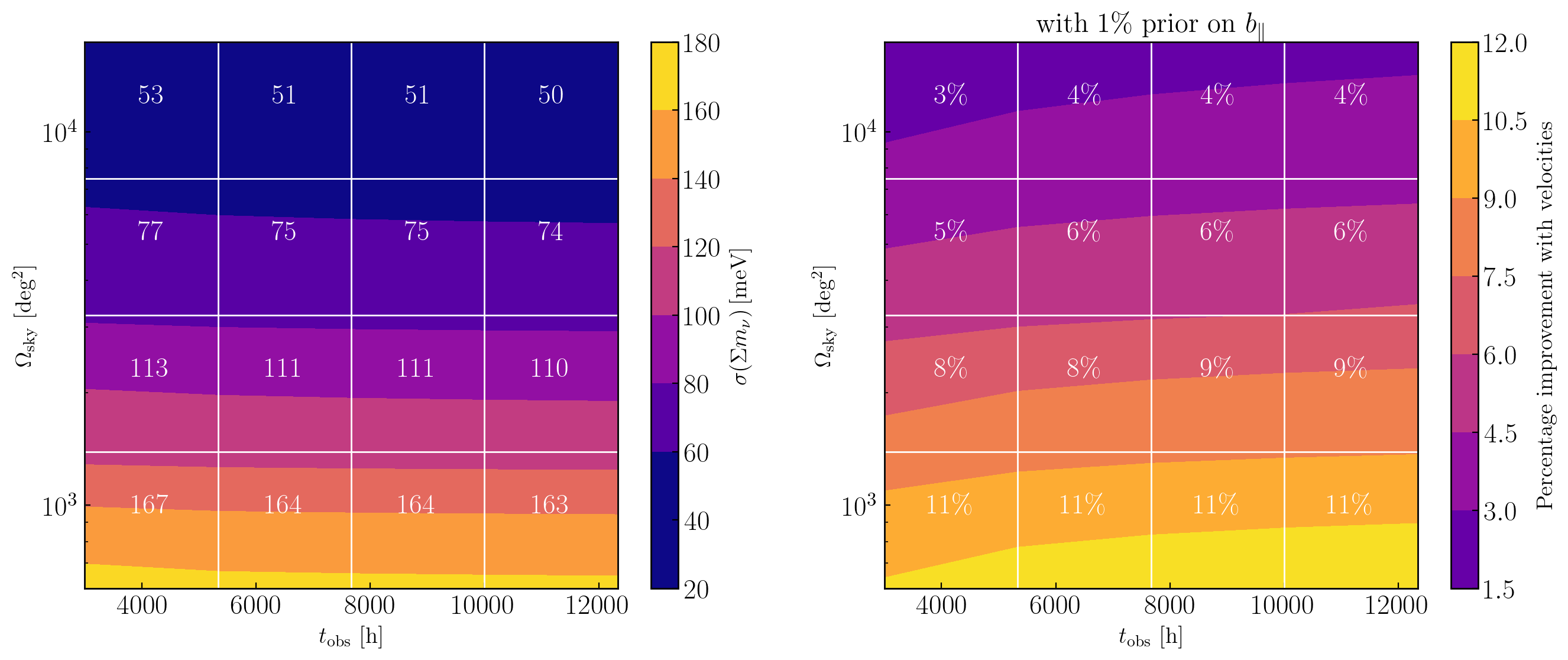}\\
	\caption{Left: Constraints on the summed neutrino masses as function of the survey area $\Omega_{\rm sky}$ and observation time $t_{\rm obs}$ of the LIM experiment. Right: Percentage improvement of the neutrino masses constraints when adding velocity information to LIM, using the full Fisher matrix from Sec.~\ref{sec:fisher}. Both panels correspond to the noise level and capability of AtLAST.}
\label{fig:lim_instrument}
\end{figure*}

Velocity tomography can further improve such constraints. 
First of all, we isolated the constraining power of the kSZ effect and the ML effect on $\sum m_\nu$, to determine the dominant factor between them. Our findings indicate that for both galaxy and LIM surveys, the kSZ effect dominates the velocity reconstruction. The ML effect, instead, only slightly contributes to the final forecast. Overall, the velocity contribution is more relevant for galaxy surveys than for LIM, due to their smaller $N_{vv}$, as seen in Fig.~\ref{fig:ksz_noise}, and to their capability of reconstructing the velocity field up to lower $z$.

The results for the full Fisher matrix slightly depend on the prior choice on the parameter $b_\parallel$ when using LIM. As Fig.~\ref{fig:sigm_mnu_bpar} illustrates, the dependence of our CII constraints with reduced-AtLAST and velocity reconstruction can reach $\sigma_{\sum {m_\nu}} \lesssim 150\,{\rm meV}$ when the prior on $b_\parallel$ is stringent enough. For galaxy surveys, instead, results are not affected by this parameter. 
{This can be understood by looking at the definition of $b_\parallel$ in Eq.~\eqref{eq:b_par}, which describes the capability of measuring the small scale correlation between density and electrons. Galaxy surveys can intrinsically observe the small scales of the density field thanks to their better angular resolution, whereas LIM window functions introduce a larger uncertainty on this parameter.}  
Since $b_\parallel$ characterizes our knowledge of the small scale cross correlation between LSS tracers and the electrons inside halos ($P_{Xe}(k,z)$, see Eq.~\eqref{eq:b_par}), its prior can be motivated by relying on external astrophysical measurements. For example, Ref.~\cite{Madhavacheril:2019buy} showed that using fast radio bursts (FRBs) observations, a 1\% measurement of $P_{Xe}$ can roughly be translated to a 1\% constraint on $b_\parallel$. To reach this level, they require to cross-correlate the dispersion measure in the time of arrival of $\mathcal{O}(10^5)$ FRB along different LoS with a galaxy survey, e.g.,~DESI; this is within reach of current and upcoming detectors on a decade timescale. Based on these results, we consider a 1\% prior on $b_\parallel$ to be a reasonable choice, and from now on we use this value to discuss our results; this, as shown in Fig.~\ref{fig:sigm_mnu_bpar}, allows us to improve $\sigma_{\sum m_\nu}$ by $\sim 10\%$ in the case of reduced-AtLAST.

We checked how the constraints change when we worsen the angular resolution $\sigma_{\rm FWHM}$ and noise level $\sigma_{\rm pix}$ of the LIM experiment (see Sec.~\ref{sec:lim_theory}). We verify that the angular resolution does not change our constraints significantly. Regarding $\sigma_{\rm pix}$, we considered increasing its value by a factor of 10 for each frequency bin, which also increases the survey sensitivity $\sigma_N$ by this factor. This enables the 1-$\sigma$ constraints  $\sigma_{\sum m_\nu}\sim 100\,{\rm meV}$ and shows a negligible improvement due to velocities.

Using a different prior for $\sum m_\nu$, e.g.,~relying on the 100\,meV case of the inverted hierarchy, has a negligible effect our conclusions, improving the constraining power of both LIM and LIM+velocities.

We can finally extend our study and explore how AtLAST constraints on $\sigma_{\sum m_\nu}$ change as a function of the observational time $t_{\rm obs}$ and the field of view $\Omega_{\rm sky}$, until we reach the final AtLAST configuration.
In the left panel of Fig.~\ref{fig:lim_instrument}, we show our constraints while varying these parameters. From the figure, we see that the larger effect is due to the field of view the detector is observing. The top right corner matches the 
AtLAST setup described in Ref.~\cite{Bernal:2020lkd}, 
where the constraints are the highest achievable, $\sigma_{\sum m_\nu} \sim 50\,{\rm meV}$. Going towards the bottom left corner we meet the reduced-AtLAST setup, forecasting constraints comparable to VRO10.

The right panel of Fig.~\ref{fig:lim_instrument} displays the percentage improvement in the constraints achieved by incorporating velocities into the LIM information. 
As we described in Sec.~\ref{sec:vel_tom}, the computation of the velocity reconstruction noise on large scales depends on our capability of reconstructing the smaller scales. In the case of LIM, this capability is limited by the noise variance per voxel: if $\sigma_N$ in Eq.~\eqref{eq:LIM_noise} gets smaller, the velocity reconstruction becomes easier. Therefore, if the observed sky area $\Omega_{\rm sky}$ remains constant and the survey duration $t_{\rm obs}$ increases (i.e.,~moving horizontally in the right panel of Fig.~\ref{fig:lim_instrument}), the larger time spent observing each voxel decreases $\sigma_N$, leading to a larger improvement due to velocities. On the other side, increasing $\Omega_{\rm sky}$ implies a larger number of voxels in the survey; if $t_{\rm obs}$ is kept constant (i.e.,~moving vertically in the plot), the noise variance increases and the relative contribution of velocities to $\Sigma m_\nu$ constraints gets weaker.
Overall, the effect of $\Omega_{\rm sky}$ is larger: as we move up in the plot, reaching the AtLAST final configuration, the velocity contribution becomes less relevant. The auto-LIM power spectrum, instead, provides direct access to the large scales when the observed sky area is large enough: for this reason, the constraining power of $P_\mathrm{XX}$ on $\Sigma m_\nu$ increases with larger $t_{\rm obs}$ and $\Omega_{\rm sky}$.

\subsection{Combined results}\label{sec:finalresults}

For a comprehensive analysis in the multi-tracer context, we study how $\sigma_{\sum m_\nu}$ improves when we incorporate information from the BAO, measured either from galaxy or LIM surveys, and CMB. To understand which are the best constraints achievable in this context, we combine the capabilities of the AtLAST detector (density and velocity tomography) with the BAO forecasts for DESI and forecasts for CMB-S4.
We calculate the Fisher matrix for CMB-S4 using the public codes \texttt{FisherLens}\footnote{\url{https://github.com/ctrendafilova/FisherLens}} and \texttt{class\_delens}~\cite{Hotinli:2021umk} for multipoles $\ell>30$. We use the Gaussian lensed covariance, and we estimate the noise using Eq.~\eqref{eq:cmb_noise} and the angular resolution and sensitivity mentioned in Sec.~\ref{sec:experiments}. 
Numerical derivatives for the Fisher matrix are computed using a 1\% step, to avoid numerical instabilities. 
For multipoles $\ell<30$, instead, we use the TT covariance matrix from Planck, which can be found in the code \texttt{pyfisher}\footnote{\url{https://github.com/msyriac/pyfisher}}. The Fisher matrix for DESI BAO is also calculated using this code. BAO forecasts could also be derived for AtLAST; Ref.~\cite{Bernal:2019gfq} showed that the constraining power of LIM experiments is comparable to the one of future galaxy surveys, hence we do not expect our results to change under this choice. 

\begin{figure} [t]
	\centering
	\includegraphics[width=\columnwidth]{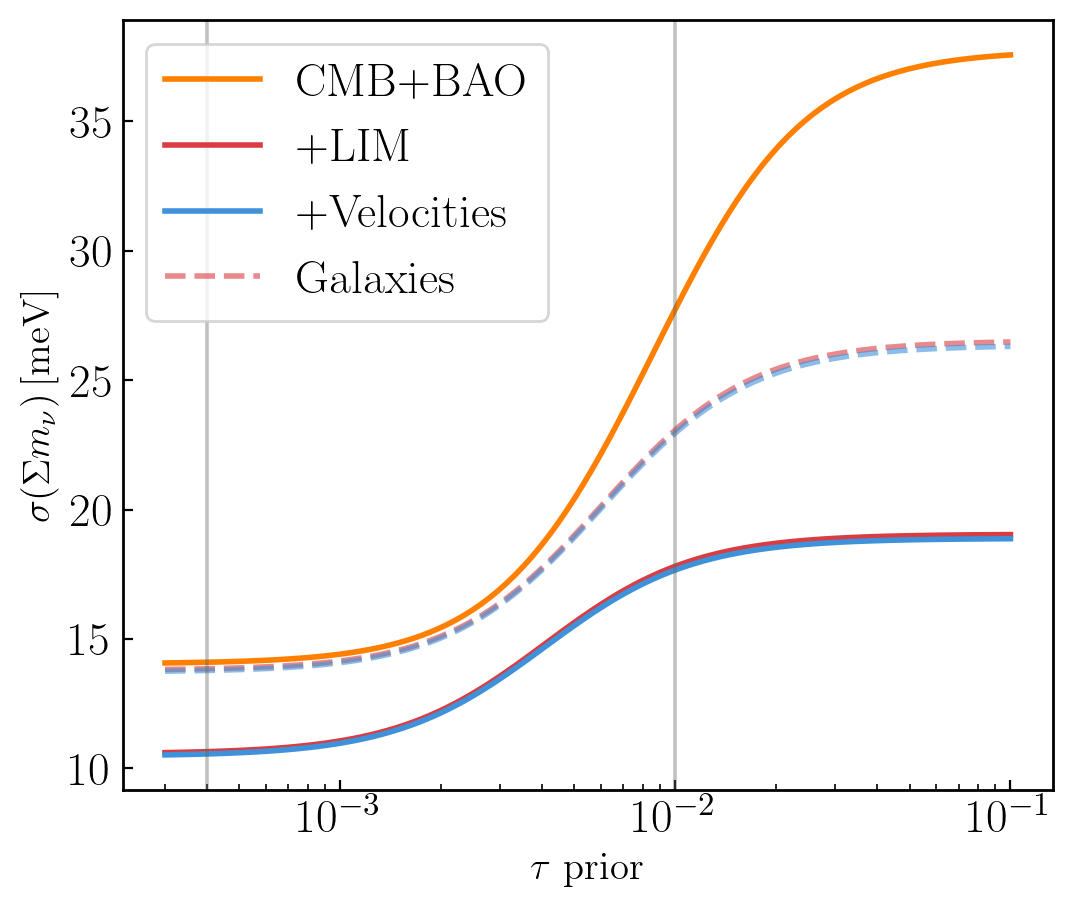}\\
	\caption{Constraints of the summed neutrino masses as function of the prior value set for $\tau$. Solid lines show constraints from LIM survey AtLAST+DESI BAO+CMB-S4 and adding velocity information, where matching colors dashed lines are from galaxy survey VRO10. The right vertical line matches the conservative prior in Ref.~\cite{CMB-S4:2016ple}, while the left vertical line indicates the prior computed in Ref.~\cite{Shmueli:2023box}, using 21-cm forecast observations.}
\label{fig:sigm_mnu_tau_+CMB+BAO}
\end{figure}

CMB-S4 constraints on $\sigma_{\sum m_\nu}$ alone suffer from degeneracy with the optical depth to reionization $\tau$~\cite{Keating:2005ds}.
Free electrons along the LoS between the observer and the surface of last scattering influence the CMB anisotropies, 
damping the scalar perturbations 
by a factor exp(-2$\tau$). This makes it highly degenerate with $A_s$, the amplitude of the primordial scalar perturbations, and at high multipoles, or smaller scales, also highly degenerate with $\Sigma m_\nu$~\cite{Reichardt:2015cos}. To overcome this issue, the CMB-S4 Fisher matrix can be summed to the Fisher for DESI-BAO and to our previously derived LIM+velocities Fisher matrix; moreover, we can study the effect of adding some $\tau$ prior to the analysis. 

The results are shown in Fig.~\ref{fig:sigm_mnu_tau_+CMB+BAO}: when combining all the surveys at our disposal, we reach constraining power of $\sigma_{\sum m_\nu} \sim$ 18 meV without the $\tau$ prior.
As anticipated and discussed earlier in this section, 
LIM inherently possesses stronger constraining power over galaxies, but for both tracers the improvement with velocity data is negligible.
This is due to the fact that velocity tomography inherits the well known degeneracies that exist between neutrino masses and cosmological parameters in the CMB, and hence it cannot be used to constrain $\sum m_\nu$ neither alone nor combined with the CMB. Using the auto-power spectrum of LIM, offers a way to break the degeneracy: when $P_\mathrm{XX}$ is combined with $P_{vv}$ and their cross-correlations (as we did in the past section) or its Fisher is summed to the CMB one, $\sigma_{\sum m_\nu}$ gains a large improvement. We note further that velocity information has a very limited constraining power by itself, much smaller than that of CMB: for this reason, their contribution in the comprehensive analysis remains subdominant.\footnote{Galaxy forecasts in Fig.~\ref{fig:sigm_mnu_tau_+CMB+BAO} are in agreement with results from Ref.~\cite{Tishue:2024}, once that small differences (e.g.,~including foregrounds in the CMB noise level, or changing the $k_S$ range) between our and their analyses are accounted for.}

Adding a $\tau$ prior can further strongly improve the constraints, {breaking some of the degeneracies with $\sum m_\nu$ in the CMB.}
The right vertical line in the plot matches the conservative prior choice in Ref.~\cite{CMB-S4:2016ple}, while the left vertical line indicates the prior computed in Ref.~\cite{Shmueli:2023box}, using 21-cm forecasts to get a direct measurement of this parameter. We see that in this regime, LIM lowers the constraints from CMB+BAO all the way to $\sigma_{\sum m_\nu} \sim$ 11 meV, which would yield a larger than $5\sigma$ measurement of $\sum m_\nu$, enabling a robust determination of the neutrino mass hierarchy problem. Folding in the 21-cm power spectrum information into the analysis will improve these constraints even further.

Alternatively, the degeneracy between $\tau$ and $\sum m_\nu$ in CMB data can be broken including the EE polarization convariance matrix in the analysis. If we rely on the EE matrix estimated from Planck low $\ell$s, the constraints become comparable to the case where a prior of $0.0075$ is used on the optical depth $\tau$.

\section{Summary and Conclusions}
\label{sec:summary}

Among the goals of modern cosmology, one of the most intriguing is to reach a measurement of the neutrino masses and solve the hierarchy problem by looking at the growth and evolution of cosmic large scale structures. 
Although galaxy surveys have been extensively studied, they come with certain limitations, such as high-cost experiments and the difficulty of detecting fainter and more distant objects. A valuable alternative that has gained more and more attention in recent years is that of line intensity mapping (LIM). In this study, we aimed to study the constraining power of LIM on neutrino masses. 
We examined the impact of massive neutrinos on various observables, arising from their free-streaming on small scales during their relativistic phase in the early Universe, causing suppression of structure formation at these scales. Given the subtlety of these effects, to reach a better and more reliable result, we explored a multi-tracer technique, in which LIM is cross correlated with CMB.
In this case, the kinetic-SZ effect and the moving-lens effect are used to reconstruct the three-dimensional velocity field, as this is sensitive to massive neutrinos through the scale-dependent growth rate and the matter power spectrum. Additionally, we incorporated data from CMB and BAO as supplementary probes for detecting massive neutrinos.

In our work, we focused on the [CII] line, but our method can be extended to other lines, e.g.,~CO or [OIII].
Our analysis demonstrates that LIM is a promising method for probing the Universe's density and velocity fields and to constrain neutrino masses. 

Future surveys will provide competitive constraints on $\Sigma m_\nu$: with a noise level as small as the one expected for the next generation detector AtLAST, an observational time of $\sim$3\,000\,hr spent on $\sim$600\,deg$^2$ will be enough to overcome the $\sigma_{\Sigma m_\nu}$ uncertainty reached by a 10 years observation with the Vera Rubin photometric galaxy survey. Velocity reconstruction obtained through cross correlation with CMB-S4 will be crucial in this case to improve the results up to $\sim 11\%$ when a $1\%$ prior on $b_\parallel$ is considered. With a larger observed sky area (7\,500\,deg$^2$) and a longer observational time (10\,000\,hr), forecasts for AtLAST indicate a sensitivity of 51 meV. Here, the relative contribution of velocity reconstruction is smaller, $\sim 4\%$, due to the larger noise per voxel. We obtained comparable results to previous studies (e.g., Refs~\cite{MoradinezhadDizgah:2021upg,Karkare:2022bai}), who derived constraints for various observational LIM setups and detectors, and explored the effect of varying the observational parameters. Further, Ref.~\cite{Dvorkin:2019jgs} suggests that our findings are in the same ballpark of forecasts for next generation CMB+LSS analyses.

Finally, we combined all the surveys at our disposal, including forecasts for CMB-S4 and DESI-BAO. When CMB and BAO are included, velocity tomography does not provide additional improvements to the constraints of LIM alone. Including the optical depth prior derived using 21-cm forecasts for HERA, leads to a sensitivity of $\sigma_{\sum m_\nu} \sim$ 11.5 meV. This would enable a $\gtrsim 5\sigma$ measurement of $\Sigma m_\nu=60$ meV.

Despite the promising results, our forecasts are for future detectors, with AtLAST and CMB-S4 not expected to be operational until the next decade or beyond. When considering current detectors, such as CCAT-prime, the constraints become significantly less stringent. Therefore, combining LIM and velocity tomography techniques with traditional galaxy surveys will remain important in the upcoming years, for advancing our understanding of neutrino masses.

Nevertheless, our findings underscore the potential of LIM and velocity tomography as complementary tools to traditional galaxy surveys for constraining neutrino masses. 
These approaches offer a pathway to more accurate cosmological measurements, paving the way for future discoveries in the field: the ability to achieve more precise measurements can significantly enhance our understanding of the properties of neutrinos and their role in the evolution of the Universe's large-scale structure.

\section*{acknowledgements}

In this work we relied on the external codes \texttt{FisherLens}, \texttt{pyfisher} (references in Sec.~\ref{sec:finalresults}) to forecast CMB-S4 and DESI-BAO.  Fig.~\ref{fig:kSZ_PS} was obtained using \texttt{tools4reionisation}
(\url{https://github.com/adeliegorce/tools4reionisation}).
Finally, LIM analysis was performed using a customized version of the \texttt{lim} code~\cite{Bernal:2019jdo,Libanore:2022ntl} (\url{https://github.com/jl-bernal/lim}). All codes are incorporated in our code (\url{https://github.com/shmuelig/LSS-Velocities}). The code is public; if you want to use it or have any questions, we are happy to share information and support.

The authors thank Gabriela Sato-Polito, Avery Tishue, Selim Hotinli and Marc Kamionkowski for useful discussions and feedback on the analysis.
GS is supported by an M.Sc.\ fellowship for female students in hi-tech fields, awarded by the Israeli Council for Higher Education. 
SL~is supported by an Azrieli International Postdoctoral Fellowship. SL thanks the John Hopkins University for the hospitality during part of the work; the visit was supported by a Balzan awarded scholarship, funded by the Balzan Centre for Cosmological Studies.
EDK acknowledges joint support from the U.S.-Israel Bi-national Science Foundation (BSF, grant No.\ 2022743) and the U.S.\ National Science Foundation (NSF, grant No.\ 2307354), and support from the ISF-NSFC joint research program (grant No.\ 3156/23).  \\

\appendix

\section{kSZ Bispectrum derivation}
\label{sec:kSZ_bispectrum}

Several statistics have been proposed to study and analyze the kSZ effect, but as was shown in Ref.~\cite{Smith:2018bpn}, they are all mathematically equivalent. The most informative in our case is the bispectrum estimator of the form $\langle \delta_X(\boldsymbol{k})\delta_X(\boldsymbol{k}') T(\boldsymbol{\ell}) \rangle$. This involves two powers of the overdensities $\delta_X$ corresponding to the LSS tracer $X$, and one power of the integrated temperature fluctuation induced by the kSZ effect, $T$. 
This three-point function can be expressed as
\begin{align}\label{eq:bispectrum1}
    &\langle \delta_X(\mathbf{k}) \delta_X(\mathbf{k'}) T(\bm{\ell}) \rangle =\\ \nonumber
    &\quad= B(\mathbf{k}, \mathbf{k'}, \bm{\ell}) (2\pi)^3 \times \delta_D^{(3)}\left(\mathbf{k}+\mathbf{k'}+\frac{\bm{\ell}}{\chi_*}\right) \\ \nonumber
    &\quad= i B(k,k',\ell,k_r)(2\pi)^3 \times\delta_D^{(3)}\left(\mathbf{k}+\mathbf{k'}+\frac{\bm{\ell}}{\chi_*}\right).
\end{align}
Since $\boldsymbol{\ell}$ is a transverse mode, the Dirac delta function implies that in the radial component $k_r + k'_r = 0$. Moreover, 2D rotational invariance implies that the bispectrum depends on the lengths of the three components only. This was captured in the second equality and the $i$ factor was introduced for convenience.

Considering the limit where all wavenumbers $k, k', \ell$ are small (e.g., large scales), we can use the approximation of the tree-level kSZ bispectrum~\cite{Smith:2018bpn}:
\begin{align}
    B(k, k', \ell, k_r) &= \frac{K_* k_r}{\chi_*^2} \left[ P_{\mathrm{X}e}(k) \frac{P_{\mathrm{X}v}(k')}{k'} -\right. \nonumber \\
    &\quad \left. - \frac{P_{\mathrm{X}v}(k)}{k} P_{\mathrm{X}e}(k') \right].
\end{align}

In our analysis in Sec.~\ref{sec:analyis}, we are interested in the squeezed limit of the bispectrum in Eq.~\eqref{eq:bispectrum1}, namely the regime in which one of the wave-vectors is much smaller than the other two. We hence distinguish between $k_L \ll k_S$, where $k_L \sim 0.1 \,\textrm{Mpc}^{-1}$ describes one of the modes associated with $\delta_X$, observed on the 
linear scale, while $k_S \sim 1 \,\textrm{Mpc}^{-1}$ describes a second mode for $\delta_X$ on the non-linear scales. The former, small wave-number, is due to the velocity field on large scales; the latter, instead, describes the small scale contribution due to the hot electron gas that sources the kSZ anisotropies. The $\ell$ mode associated with the CMB temperature fluctuations describes as well a long mode; since kSZ contributions are non-negligible at  $\ell \sim 4000$,
the delta function in Eq~\eqref{eq:bispectrum1} forces the large wave-number to be around $1 \, \textrm{Mpc}^{-1}$. Note that, with the new notation the triangle condition (set by the delta function) $\bm{k}_L + \bm{k}_S + (\bm{\ell} / \chi_*) = 0$ implies $\ell \approx (k_S \chi_*)$ and $k_{S_r} = - k_{L_r}$. The equality between the radial components translates to a relation between the angles, namely $\mu_S = \mu_L k_L/k_S$, where we re-absorbed the minus sign that relates $k_S$ and $k_L$ into $\mu_L$.
 From here, we get the final bispectrum expression:
\begin{equation}
    B(k_L,k_S,\ell,\mu_L) = \frac{K_* \mu_L}{\chi_*^2} P_{\mathrm{X}v}(k_L) P_{\mathrm{X}e}(k_S),
\end{equation}
where we neglected the second term of the expression in Eq.~\eqref{eq:bispectrum1}, this being smaller than the first term.

\section{Derivation of $N_{\Phi\Phi}$}
\label{sec:N_phiphi}

Let us derive the expression for the noise for the gravitational potential power spectrum $N_{\Phi\Phi}$ used in Sec.~\ref{sec:ML_tom}, when this is estimated by studying a generic LSS tracer $X$.
The $X$ auto-power spectrum is of the form
\begin{equation}
    P_{\mathrm{XX}} = b_{X}^2(k) P_m
\end{equation}
where $b_{X}(k)$ is the tracer clustering bias, and $P_m$ is the matter power spectrum. 
The matter density field can be related to the gravitational potential via the Poisson equation in expanding space,
\begin{align}
    \nabla^2 \Phi &= 4\pi G \rho_m a^2 \delta_{m}, \\
    \delta_{m}(k) &= \frac{k^2}{4\pi G \rho_m a^2} \Phi(k).
\end{align}
where $G$ is the gravitational constant, $\rho_m$ in the matter density, $a$ is the scale factor, and in the second line we switch to Fourier space. From here, we get the relation
\begin{align}
    P_{\mathrm{XX}} = b_{X}^2(k) \left( \frac{k^2}{4\pi G \rho_m a^2} \right)^2 P_{\Phi\Phi},
\end{align}
where, using the relations $\rho_m = (1+z)^3 \rho_{m,0}$, 
$\rho_{m,0} = {3 H_0^2 \Omega_m}/({8\pi G})$
and $a = ({1+z})^{-1}$, we have
\begin{align}
    P_{\mathrm{XX}} = b_{X}^2(k) \left( \frac{2k^2}{(1+z)\rho_{m,0}} \right)^2 P_{\Phi\Phi}.
\end{align}
We can finally invert this expression to obtain the noise term
\begin{equation}
    N_{\Phi\Phi} = \left[ \frac{(1+z) \rho_{m,0}}{2} \right]^2 \frac{1}{k^4 b_{X}^2(k)} N_{\mathrm{XX}}.
\end{equation}

\end{document}